\documentclass[prb,twocolumn,eqsecnum,showpacs]{revtex4}

\usepackage{graphicx}

\begin{document}  

\title{Structure and the Lamb-shift-like quantum splitting\\
 of the pseudo-zero-mode 
Landau levels in bilayer graphene
}
\author{K. Shizuya}
\affiliation{Yukawa Institute for Theoretical Physics\\
Kyoto University,~Kyoto 606-8502,~Japan }

\begin{abstract}

In a magnetic field bilayer graphene supports an octet of 
zero-energy Landau levels with an extra twofold degeneracy 
in Landau orbitals $n=0$ and $n=1$.
It is shown that this orbital degeneracy is lifted 
due to Coulombic quantum fluctuations of the valence band (the Dirac sea); 
this is a quantum effect analogous to the Lamb shift in the hydrogen atom.  
A detailed study is made of how these zero-energy levels evolve, with filling, 
into a variety of pseudo-zero-mode Landau levels 
in the presence of possible spin and valley breaking 
and Coulomb interactions, 
and a comparison is made with experimental results.

\end{abstract} 


\maketitle


\section{Introduction}

Graphene,~\cite{NG,ZTSK,ZA,PGN,AF}
 an atomic layer of graphite, supports as charge carriers 
Dirac fermions
that lead to unique and promising electronic properties.
Of particular interest recently are bilayers~\cite{NMMKF,MF} 
(and multilayers) of graphene, 
where an added layer degree of freedom combines 
with spin and valley degrees 
to make the physics of graphene far richer.
In particular, bilayer graphene has 
the property that its band gap is 
externally controllable.~\cite{MF,OBSHR,Mc,CNMPL,OHL}

Quantum phenomena associated with Dirac fermions become
prominent in a magnetic field.
In a magnetic field graphene supports a set of characteristic 
zero-energy Landau levels, whose emergence and degeneracy 
have a topological origin in the nonzero index of the Dirac operators, 
or in the chiral anomaly in 1+1 dimensions.~\cite{NS}
Monolayer graphene has four such zero-energy levels owing to 
the spin and valley degeneracy.

Bilayer graphene supports an octet of such zero-energy levels,
due to an extra twofold degeneracy~\cite{MF} in Landau orbitals  $n=0$ and $n=1$.
This orbital degeneracy brings about a new realm of 
quantum phenomena,~\cite{BCNM,KSpzm,BCLM,CLBM,CLPBM,CFL}
such as orbital mixing  and orbital-pseudospin waves.
The eightfold degeneracy of the zero-energy levels 
is partially or fully lifted 
in the presence of Zeeman coupling, 
interlayer bias and Coulomb interactions, and 
these levels evolve into a variety of pseudo-zero-mode levels, 
or broken-symmetry states, as discussed theoretically
in the context of quantum-Hall ferromagnetism~\cite{BCNM}
and others.~\cite{NL,GGJM}
Experimentally the full splitting of the pseudo-zero-mode levels 
has been observed~\cite{FMY,ZCZJ,WAFM,VJB}
in bilayer graphene.

A key feature of graphene is that 
graphene is an intrinsically many-body system of electrons equipped 
with the valence band that acts as the Dirac sea.
Quantum fluctuations of the Dirac sea are sizable 
and lead to such quantum phenomena as 
velocity renormalization,~\cite{velrenorm}
screening of charge,~\cite{KSbgr} 
and nontrivial Coulombic corrections to cyclotron resonance.\cite{JHT,IWFB,BMgr,KCC,KScr}
It will be important to ask how such quantum fluctuations affect 
the pseudo-zero-mode sector in bilayer graphene, in order to 
interpret experimental results on broken-symmetry states properly,

The purpose of this paper is to study the structure and spectra
of the pseudo-zero-mode sector in bilayer graphene.
It is shown, in particular, that the orbital degeneracy of the zero-energy levels 
is lifted by Coulombic vacuum fluctuations, 
leading to an appreciable shift and splitting 
of the $n=0$ and $n=1$ levels; this is 
a quantum effect analogous to the Lamb shift\cite{Lambshift,KKS} 
in the hydrogen atom.
This vacuum effect is correlated with the Coulomb interaction 
within the pseudo-zero-mode octet
to yield eventually a particle-hole symmetric spectrum for this octet.
Unexpectedly, a detailed analysis of the interlayer Coulomb interaction reveals 
negative capacitance energies for bilayer graphene, 
which suppress possible valley rotations in the pseudo-zero-mode sector.
We discuss how this  sector is split in spin, valley and orbital,
with filling, in the presence of Zeeman coupling, interlayer bias 
and Coulomb interactions,
and compare with experimental results.

In Sec.~II we briefly review some basic features of 
the pseudo-zero-mode octet in bilayer graphene, 
and in Sec.~III show that the orbital degeneracy of this octet is lifted 
due to quantum fluctuations.
In Sec.~IV  we discuss in a simplified setting how this quantum effect cooperates
with the Coulomb exchange interaction to determine the spectra 
of the  pseudo-zero-mode levels.
In Sec.~V we take a close look into the valley breaking due to the interlayer Coulomb interaction.
In Sec.~VI we examine the hierarchy of broken-symmetry states 
under general and practical conditions 
with spin, valley and orbital breakings, and in Sec.~VII compare 
with experimental results. 
Section~VIII is devoted to a summary and discussion.

\section{bilayer graphene}

Bilayer graphene consists of two coupled honeycomb lattices of 
carbon atoms in Bernal $A'B$ stacking.
The electrons in it  are described 
by four-component spinor fields on the four inequivalent sites 
$(A,B)$ and $(A',B')$ in the bottom and top layers,
and their low-energy features are governed 
by the two inequivalent Fermi points $K$ and $K'$ in the Brillouin zone. 
The intralayer coupling
$\gamma_{0} \equiv \gamma_{AB} \sim 3$\, eV
is related to the Fermi velocity 
$v = (\sqrt{3}/2)\, a_{\rm L}\gamma_{0}/\hbar \sim 10^{6}$~m/s 
(with $a_{\rm L}=  0.246$nm) in monolayer graphene.
Interlayer hopping via the $(A',B)$ dimer coupling~\cite{Malard} 
$\gamma_{1} \equiv \gamma_{A'B} \sim 0.4$\,eV 
modifies the intralayer linear spectra
to yield quasi-parabolic spectra~\cite{MF} 
in the low-energy branches $|\epsilon| <\gamma_{1}$.

The effective Hamiltonian with such leading intra- and inter-layer couplings 
is written as~\cite{MF}
\begin{eqnarray}
H^{\rm bi} &=&\!\! \int\! d^{2}{\bf x}\, \Big[ (\Psi^{K})^{\dag}\, {\cal H}_{K} \Psi^{K}
+ (\Psi^{K'})^{\dag}\, {\cal H}_{K'}\, \Psi^{K'}\Big], \nonumber\\
{\cal H}_{K} &=& \left(
\begin{array}{cccc}
{1\over{2}}u &  &   & v\,p^{\dag} \\
  & -{1\over{2}}u & v\,p &  \\
 & v\,  p^{\dag} &   -{1\over{2}}u & \gamma_{1} \\
v\,p & & \gamma_{1}  & {1\over{2}}u \\
\end{array}
\right),
\label{Hbilayer}
\end{eqnarray}
with $p= p_{x}+ i\, p_{y}$, $p^{\dag}= p_{x} - i\, p_{y}$.
Here $\Psi^{K} = (\psi_{A},\psi_{B'},\psi_{A'}, \psi_{B})^{\rm t}$
stands for the electron field at the $K$ valley, with $A$ and $B$ 
referring to the associated sublattices; 
$u$ stands for the interlayer bias, 
which opens a tunable gap~\cite{Mc,CNMPL} 
between the $K$ and $K'$ valleys.
We ignore the effect of trigonal warping 
$\propto \gamma_{3}\equiv \gamma_{AB'} \sim 0.1$\,eV 
which, in a strong magnetic field, causes 
only a negligibly small level shift~\cite{KSbgr};
also ignored are some nonleading intra- and inter-layer couplings 
that lead to weak electron-hole asymmetry~\cite{ehasymmetry} 
in bilayer graphene. 
${\cal H}_{K}$ is diagonal in the (suppressed) electron spin.

The Hamiltonian ${\cal H}_{K'}$ at another valley is given by ${\cal H}_{K}$ 
with $(v,u) \rightarrow (-v, -u)$, 
and acts on a spinor of the form 
$\Psi^{K'} = (\psi_{B'},\psi_{A},\psi_{B}, \psi_{A'})^{\rm t}$.
Actually, ${\cal H}_{K'}$ is unitarily equivalent to 
${\cal H}_{K}$ with the sign of $u$ reversed,  
\begin{equation}
{\cal H}_{K'}|_{u} =S^{\dag}{\cal H}_{K}|_{-u}\, S
\label{Hequi}
\end{equation}
with $S= {\rm diag}(1,1,-1,-1)$.
In what follows we adopt ${\cal H}_{K}|_{-u}$ for ${\cal H}_{K'}$
and simply pass to the $K'$ valley by reversing the sign of $u$ 
in the $K$-valley expressions.
Nonzero bias $u\not=0$ thus acts as a valley-symmetry breaking.

Let us place bilayer graphene in a strong uniform magnetic field 
$B_{z} = -B<0$ normal to the sample plane;
we set, in ${\cal H}_{K}$, 
$p\rightarrow \Pi = p + eA$
with $A= A_{x}+ iA_{y}= B\, y$, 
and denote the the magnetic length as $\ell=1/\sqrt{eB}$.
It is easily seen that the eigenmodes of ${\cal H}_{K}$ have the structure 
\begin{equation}
\Psi_{n} = \Big(|n\rangle\, b_{n}^{(1)} ,|n\!-\!2\rangle\, b_{n}^{(2)},
|n\!-\!1\rangle\, b_{n}^{(3)}, |n\!-\!1\rangle\, b_{n}^{(4)}\Big)^{\rm t}
\end{equation} 
with $n=0,1,2,\cdots$, where only the orbital eigenmodes are shown 
using the standard harmonic-oscillator basis $\{ |n\rangle \}$
(with the understanding that $|n\rangle =0$ for $n<0$).
The coefficients 
${\bf b}_{n}=(b_{n}^{(1)}, b_{n}^{(2)}, b_{n}^{(3)}, b_{n}^{(4)})^{\rm t}$
 for $n=2,3,\dots$ are given by the eigenvectors of the reduced Hamiltonian
\begin{equation}
\hat{\cal H}_{\rm red} = \omega_{c}\, \left(
\begin{array}{cccc}
M &   &  & \sqrt{n} \\
 & - M & \sqrt{n-1} & \\
 & \sqrt{n-1}  &  - M & \gamma' \\
\sqrt{n} &  & \gamma'  & M \\
\end{array}
\right),
\label{reducedH}
\end{equation} 
where 
\begin{equation}
\omega_{c}\equiv \sqrt{2}\, v/\ell 
\approx 36.3 \times v[10^{6}{\rm m/s}]\, \sqrt{B[{\rm T}]}\ {\rm meV},
\end{equation} 
with $v$ measured in units of $10^{6}$m/s and $B$ in tesla, 
is the characteristic cyclotron energy 
for monolayer graphene;
$M\equiv {1\over{2}}\, u/\omega_{c}$ and  $\gamma'\equiv \gamma_{1}/\omega_{c}$.

The energy eigenvalues obey the secular equation
\begin{eqnarray}
&& (\gamma')^{2}\,(\epsilon'^{2} -M^{2})
\nonumber\\
&&=\{ (\epsilon'-M)^{2}-|n|\}\,
 \{ (\epsilon'+M)^{2}-|n|+1\},
\label{seceq}
\end{eqnarray}
where $\epsilon' \equiv \epsilon_{n}/\omega_{c}$.
Let us  denote the four solutions of the secular equation as
$\epsilon_{-n}^{-} < \epsilon_{-n}<0<\epsilon_{n}
< \epsilon_{n}^{+}$ for each integer $n\ge 2$ and $|M|<1$, 
so that the index $\pm n$ reflects the sign of $\epsilon_{n}$. 
The eigenvectors ${\bf b}_{n}$ for $|n|\ge 2$ are written as
\begin{eqnarray}
b^{(2)}_{n}&=&  \{\sqrt{|n|-1}/(\epsilon' +M) \} \,b^{(3)}_{n}, 
\nonumber\\
b^{(3)}_{n}/b^{(1)}_{n} &=& - (1/\gamma')\, \{|n| - (\epsilon' -M)^{2}\}/\sqrt{|n|}, 
\nonumber\\
b^{(4)}_{n}/b^{(1)}_{n} &=& (\epsilon' -M)/\sqrt{|n|},
\end{eqnarray}
with $b^{(1)}_{n}$ fixed by normalization $|{\bf b}_{n}| =1$.
These expressions are equally valid for 
$\epsilon_{n} \rightarrow  \epsilon_{n}^{\pm}$.

Of our particular concern are 
the $n=0$ and $n=1$ modes.
For $n=0$, $\hat{\cal H}_{\rm red}$ has an obvious eigenvalue 
$\epsilon_{0}={1\over{2}} u \equiv \omega_{c}\, M$ with eigenvector
${\bf b}_{0} = (1,0,0,0)^{\rm t}$.
For $n=1$ Eq.~(\ref{seceq}) has three solutions
$(\epsilon^{-}_{-1}, \epsilon_{1},\epsilon^{+}_{1})$
(excluding $\epsilon'=-M$).
Actually $|\epsilon^{\pm}_{\pm1}| >\gamma_{1}$ and we focus on $\epsilon_{1}$ 
which is close to $\epsilon_{0}$. Let us set $\epsilon_{1}= {1\over{2}}u \, (1-z)$,
with $z <1$ determined from 
\begin{equation}
z = {1\over{(\gamma'})^{2}}(2-z)(1- M^{2}z^{2})
\stackrel{M\rightarrow 0}{=} {2\over{(\gamma')^{2}+1}}.
\end{equation}
The associated eigenvector ${\bf b}_{1}$ 
is written as
\begin{eqnarray}
{\bf b}_{1}&=& c_{1}\, \Big(1,0,-{1\over{\gamma'}}(1- M^{2}z^{2}), -zM\Big),
\nonumber\\
c_{1}^{2}&=&1/\Big[1+ {1\over{\gamma'^{2}}} (1- M^2 z^2)^2 + z^2 M^2\Big]
\equiv 1- {{z}_{c}\over{2}}. \ \ \ \ \ 
\label{beone}
\end{eqnarray}
Both $z$ and $z_{c}$, defined above, are functions of $(M^2, \gamma'^{2})$, 
and are thus common to the $K$ and $K'$ valleys. 
They coincide for $M=0$, $z = {z}_{c} =2/({\gamma'}^{2}+1)$;
numerically, $z\approx 0.181$ and $c_{1}^{2} \approx 0.909$
for $M=0$ at $B$=10 T, with $v \approx 1.1 \times 10^{6}$ m/s
and $\gamma_{1} \approx 0.4$ eV.
These $\epsilon_{0}$ and $\epsilon_{1}$ modes are called 
the pseudo-zero-modes
since they evolve from the zero-energy modes of the $M=0$ case.
For $M=0$ there are eight such zero-energy Landau levels 
differing in spin, valley and orbital $[n=(0,1)]$ degrees of freedom;
the presence of the zero-energy modes is dictated 
by the nonzero index~\cite{NS, KSbgr} 
of the Dirac Hamiltonian ${\cal H}_{K} \oplus {\cal H}_{K'}$.

One can pass to the $K'$ valley by setting $M\rightarrow -M$
in the $K$ valley expressions.
The spectra $(\epsilon_{0}, \epsilon_{1})$ 
thereby change sign but 
for later convenience we continue to use $n=(0,1)$ 
to specify the pseudo-zero-mode levels at the $K'$ valley;
we thus write $\epsilon_{n}|_{K'} = -\epsilon_{n}|_{K}$
for $n\in (0,1)$.
The eigensystems at the two valleys are related as
\begin{eqnarray}
&&(b_{n}^{(1)}, b_{n}^{(2)}, b_{n}^{(3)}, b_{n}^{(4)})|_{K'} 
= (b_{-n}^{(1)}, - b_{-n}^{(2)}, b_{-n}^{(3)}, -b_{-n}^{(4)})|_{K}, 
\nonumber\\
&&\epsilon_{n}|_{K'} = -\epsilon_{-n}|_{K},
\label{bKtoK}
\end{eqnarray}
for $|n| \ge 2$ [and for $n\in (0,1)$ as well if one sets 
$\pm 0\rightarrow 0$
and $\pm1\rightarrow 1$].

The Landau-level structure is made explicit by passing to
the $|n,y_{0}\rangle$ basis  (with $y_{0}\equiv \ell^{2}p_{x}$) 
via the expansion
$(\Psi^{K} ({\bf x}), \Psi^{K'} ({\bf x}) ) 
= \sum_{n, y_{0}} \langle {\bf x}| n, y_{0}\rangle\, \{ \psi^{n;a}_{\alpha}(y_{0})\}$, 
where $n$ refers to the Landau level index, 
$\alpha \in (\uparrow, \downarrow)$ 
to the electron spin and 
$a \in (K,K')$ to the valley.
The charge density 
$\rho_{-{\bf p}} =\int d^{2}{\bf x}\,  e^{i {\bf p\cdot x}}\,\rho$ 
with $\rho = (\Psi^{K})^{\dag}\Psi^{K} +  (\Psi^{K'})^{\dag}\Psi^{K'}$ 
is thereby written as~\cite{KSbgr}
\begin{eqnarray}
\rho_{-{\bf p}} &=& \gamma_{\bf p}\sum_{k, n =-\infty}^{\infty}
\sum_{a,\alpha} g^{k n;a}_{\bf p}\, 
R^{k n;aa}_{\alpha\alpha;\bf p}, \nonumber\\
R^{kn;ab}_{\alpha\beta;{\bf p}}&\equiv& \int dy_{0}\,
{\psi^{k;a}_{\alpha}}^{\dag}(y_{0})\, e^{i{\bf p\cdot r}}\,
\psi^{n;b}_{\beta} (y_{0}),
\label{chargeoperator}
\end{eqnarray}
where $\gamma_{\bf p} =  e^{- \ell^{2} {\bf p}^{2}/4}$; 
${\bf r} = (i\ell^{2}\partial/\partial y_{0}, y_{0})$
stands for the center coordinate with uncertainty 
$[r_{x}, r_{y}] =i\ell^{2}$.
In particular, the charge operators 
$\sum_{a,\alpha} R^{k n;aa}_{\alpha\alpha;\bf p}$ obey 
the $W_{\infty}$ algebras~\cite{GMP}
associated with intralevel center-motion 
and interlevel mixing of electrons.

The coefficient matrix $ g^{k n;a}_{\bf p} \equiv g^{kn}_{\bf p}|_{a}$ 
at valley $a\in (K,K')$ is constructed 
from the knowledge of the eigenvectors ${\bf b}_{n}|_{a}$,
\begin{eqnarray}
g^{kn}_{\bf p} &=& b_{k}^{(1)}\, b_{n}^{(1)}\, f_{\bf p}^{|k|,|n|}
+ b_{k}^{(2)}\, b_{n}^{(2)}\, f_{\bf p}^{|k|-2,|n|-2} \nonumber\\
&&+ (b_{k}^{(3)}\, b_{n}^{(3)}+ b_{k}^{(4)}\, b_{n}^{(4)})\, f_{\bf p}^{|k|-1,|n|-1},
\label{gkn}
\end{eqnarray}
where
\begin{equation}
f^{k n}_{\bf p} 
= \sqrt{{n!\over{k!}}}\,
\Big({-\ell p\over{\sqrt{2}}}\Big)^{k-n}\, L^{(k-n)}_{n}
\Big ({1\over{2}} \ell^{2}{\bf p}^{2}\Big)
\label{fknp}
\end{equation}
for $k \ge n\ge0$, and $f^{n k}_{\bf p} = (f^{k n}_{\bf -p})^{\dag}$;
$p=p_{x}\! +i\, p_{y}$; it is understood that 
$f^{kn}_{\bf p}=0$ for $k<0$ or $n<0$.
As seen from Eq.~(\ref{bKtoK}), they are related at the two valleys so that 
\begin{equation}
g^{mn;K'}_{\bf p}=g^{-m,-n;K}_{\bf p},\ \ 
g^{mn;a}_{\bf p}|_{M}=g^{-m,-n;a}_{\bf p}|_{-M},
\label{gmnproperty}
\end{equation}
where it is understood that one sets $\pm 1\rightarrow 1$ and $\pm0 \rightarrow 0$.
Within the $n\in (0,1)$ sector, they are common to both valleys, 
\begin{eqnarray}
g^{00}_{\bf p} &=& 1, \ \
g^{11}_{\bf p} =1-c_{1}^{2}\, \textstyle{1\over{2}}\ell^{2}{\bf p}^{2}, 
\nonumber\\
g^{01}_{\bf p} &=& ic_{1} \ell\, p^{\dag}/\sqrt{2},\ \ 
g^{10}_{\bf p} = ic_{1} \ell\, p/\sqrt{2},
\end{eqnarray}
so that the valley index $a$ may be suppressed.

From now on we frequently suppress
summations over levels $n$, spins $\alpha$ and valleys $a$, 
with the convention that the sum is taken over repeated indices.
The Hamiltonian $H^{\rm bi}$ projected to the octet of 
pseudo-zero-mode Landau levels is thereby written as
\begin{equation}
H_{u} = \epsilon_{u}^{n}\, \delta R^{nn}_{\beta\beta;{\bf 0}} 
- \mu_{\rm Z}\, (T_{3})_{\beta\alpha} R^{nn;aa}_{\alpha\beta;{\bf 0}} 
\label{Hzero}
\end{equation}
with $n\in (0,1)$, where $\epsilon_{u}^{0} = {1\over{2}}\, u$ and 
$\epsilon_{u}^{1} ={1\over{2}}\, u\,  (1 - z)$;
$\delta R^{mn}_{\alpha\beta;{\bf 0}} \equiv R^{mn;KK}_{\alpha\beta;{\bf 0}}  
-R^{mn;K'K'}_{\alpha\beta;{\bf 0}}$. 
Here the Zeeman term $\mu_{\rm Z} \equiv g^{*}\mu_{\rm B}B$ is introduced 
via the spin matrix 
$T_{3} = \sigma_{3}/2$,
As interlayer bias $u$ is turned on, these $n=(0,1)$ levels go up or down oppositely 
at the two valleys.
Nonzero $u$ thus  critically breaks  the valley symmetry of the $n=(0,1)$ sector.

\section{vacuum fluctuations}

The Coulomb interaction is written as
\begin{equation}
V = {1\over{2}} \sum_{\bf p}
v_{\bf p}\, :\rho_{\bf -p}\, \rho_{\bf p}:,
\label{Hcoul}
\end{equation}
where
$v_{\bf p}= 2\pi \alpha/(\epsilon_{\rm b} |{\bf p}|)$ is 
the Coulomb potential with 
$\alpha = e^{2}/(4 \pi \epsilon_{0}) \approx 1/137$ and 
the substrate dielectric constant $\epsilon_{\rm b}$;
$\sum_{\bf p} =\int d^{2}{\bf p}/(2\pi)^{2}$.   
In this section we show that the orbital degeneracy of the pseudo-zero-mode 
octet is lifted by Coulombic quantum fluctuations 
of the valence band (the Dirac sea).

Let us define the Dirac sea $|{\rm DS}\rangle$ as the valence band 
with levels with $n\le -2$ are all filled.
We take the expectation value $\langle {\rm DS}|V|{\rm DS}\rangle$ 
to construct the effective Hamiltonian that governs the electron states 
over $|{\rm DS}\rangle$.
The best way to achieve this is to construct the Hartree-Fock (HF) Hamiltonian
$V^{\rm HF}$ out of $V$.

In this paper we generally focus on many-body ground states $|G\rangle$ 
with a homogeneous density, realized at integer filling factor $\nu \in [-4, 4]$,
and set the expectation values  
$\langle G| R^{mn;ab}_{\alpha\beta; {\bf k}}|G \rangle 
= \delta_{\bf k,0}\, \rho_{0}\,  \nu^{mn;ab}_{\alpha\beta}$
with $\rho_{0} = 1/(2\pi \ell^{2})$ and $\delta_{\bf k,0}= (2\pi)^2\, \delta^2({\bf k})$.
Accordingly, the filling factor $\nu^{nn;aa}_{\alpha\alpha}=1$ 
when the Landau level specified by 
$(n,a,\alpha)$ is filled up.

Let us divide $V^{\rm HF}$ into two parts, $V^{\rm HF} = V_{\rm D}+ V_{\rm X}$. 
The direct interaction reads
\begin{equation}
V_{\rm D}
=\rho_{0}v_{\bf p\rightarrow 0}\,
\nu^{nn;aa}_{\alpha\alpha}\,  R^{m'm';bb}_{\beta\beta; {\bf 0}}
\end{equation}
while the exchange interaction is written as
\begin{equation}
V_{\rm X}
= - \sum_{\bf p}v_{\bf p}\gamma_{\bf p}^{2}\,
g^{m n';b}_{\bf -p}\,g^{m'n;a}_{\bf p}\,
\nu^{mn;ba}_{\beta \alpha}\, R^{m' n';ab}_{\alpha\beta;{\bf 0}};
\label{Vex}
\end{equation}
note that $g^{mn;a}_{{\bf p}=0}=\delta^{mn}$.
In $V_{\rm D}$ and $V_{\rm X}$ we sum over filled levels $(m,n)$
and retain the pseudo-zero-mode sector $m',n'\in (0,1)$.
As usual,  the direct term $V_{\rm D}$ with a divergent factor $v_{\bf p\rightarrow 0}$
is removed if one takes into account neutralizing  uniform
positive background charges.

Let us first extract, out of $V_{\rm X}$,
the contribution from the Dirac sea, i.e., all filled levels with $n\le -2$,
\begin{equation}
V^{\rm DS}_{\rm X}
= - \sum_{\bf p}v_{\bf p}\gamma_{\bf p}^{2}\,
\sum_{n\le -2}|g^{m'n;a}_{\bf p}|^2\, 
R^{m' m';aa}_{\alpha\alpha;{\bf 0}},
\label{VxDS}
\end{equation}
where the sum over $m'\in (0,1)$, $a\in (K,K')$ 
and $\alpha \in (\uparrow, \downarrow)$ is understood.
Actually, the sum over infinitely many filled levels 
with $-N_{L} < n\le -2$
gives rise to  an ultraviolet divergence  with $N_{L}\rightarrow \infty$.

Fortunately one can isolate the divergence 
and even evaluate $V^{\rm DS}_{\rm X}$ exactly for $M\rightarrow 0$, as shown below.
It is clear from Eq.~(\ref{gmnproperty}) that 
$g^{mn;K}_{\bf p} = g^{mn;K'}_{\bf p}=g^{-m,-n}_{\bf p}$ for $M=0$.
For $M\rightarrow 0$ one can thus replace the sum
$\sum_{n\le -2} \rightarrow $ ${1\over{2}}\, (\sum_{n\le -2} + \sum_{n\ge2})$
in Eq.~(\ref{VxDS}) 
and note the completeness relation~\cite{fn} 
\begin{equation}
\sum_{n=-\infty}^{\infty} |g^{mn}_{\bf p}|^{2} = e^{\ell^{2}{\bf p }^{2}/2}
\label{comp}
\end{equation}
to obtain
\begin{eqnarray}
\sum_{n\le -2} |g^{0n}_{\bf p}|^{2} 
&\stackrel{M=0}{=}& {1\over{2}}\, (e^{\ell^{2}{\bf p }^{2}/2} - |g^{00}_{\bf p}|^{2}
- |g^{01}_{\bf p}|^{2}),
\nonumber\\
\sum_{n\le -2} |g^{1n}_{\bf p}|^{2} &\stackrel{M=0}{=}& 
 {1\over{2}}\, (e^{ \ell^{2}{\bf p }^{2}/2} - |g^{10}_{\bf p}|^{2}
- |g^{11}_{\bf p}|^{2}).
\label{DSsum}
\end{eqnarray}
We have independently confirmed this result by direct numerical calculations.

The $e^{ \ell^{2}{\bf p }^{2}/2}$ term in Eq.~(\ref{DSsum}) 
leads to a divergence upon integration over ${\bf p}$;
it is, however, common to all levels $m'$ and is safely omitted.
We thus take $-{1\over{2}}\, (|g^{00}_{\bf p}|^{2} + |g^{01}_{\bf p}|^{2})$ as
the regularized expression for $\sum_{n\le -2} |g^{0n}_{\bf p}|^{2}$; 
and $-{1\over{2}}\, (|g^{11}_{\bf p}|^{2} + |g^{10}_{\bf p}|^{2}) $ 
for $\sum_{n\le -2} |g^{1n}_{\bf p}|^{2}$.
One can make use of these $M=0$ expressions 
to evaluate the $M\not=0$ corrections unambiguously.

 Substituting these regularized expressions into  Eq.~(\ref{VxDS}) 
 and integrating over ${\bf p}$ yield
\begin{equation}
V^{\rm DS}_{\rm X}
\stackrel{M\rightarrow 0}{=} \epsilon_{\rm v}^{0}\, R^{00;aa}_{\alpha\alpha;{\bf 0}}
+ \epsilon_{\rm v}^{1}\, R^{11;aa}_{\alpha\alpha;{\bf 0}},
\label{VxDStwo}
\end{equation}
with
\begin{eqnarray}
\epsilon_{\rm v}^{0} &=& \sum_{\bf p}v_{\bf p}\gamma_{\bf p}^{2}\, 
\textstyle{1\over{2}} (|g^{00}_{\bf p}|^{2} + |g^{01}_{\bf p}|^{2}) 
= {1\over{2}}\, (1+ {1\over{2}}c_{1}^{2})\, \tilde{V}_{c},
\nonumber\\
\epsilon_{\rm v}^{1} &=& \sum_{\bf p}v_{\bf p}\gamma_{\bf p}^{2}\, 
\textstyle{1\over{2}}(|g^{11}_{\bf p}|^{2} + |g^{10}_{\bf p}|^{2})
= \textstyle{1\over{2}} (1+ {1\over{2}} c_{1}^{2} -\! {1\over{4}} C )\,  \tilde{V}_{c},
\nonumber\\
C&=&  c_{1}^{2}\, (4 - 3\,  c_{1}^{2}),
\label{Vzeroreg}
\end{eqnarray}
where 
\begin{eqnarray}
&&\sum_{\bf p}v_{\bf p}\gamma_{\bf p}^{2} 
= \sqrt{\pi/2}\, V_{c} \equiv  \tilde{V}_{c},
\nonumber\\
&&V_{c} \equiv \alpha/(\epsilon_{b}\ell) \approx (56.1/\epsilon_{b})\, 
\sqrt{B[{\rm T}]}\, {\rm meV}.
\end{eqnarray}
Numerically, for $\gamma_{1} = 0.4$ eV, $v=1.1 \times 10^6$ m/s  and $M=0$, 
one has
$c_{1}^{2} =0.909$ and $C= 1.156$ at $B=10$T, which give
$\epsilon_{\rm v}^{0} \approx 0.73\, \tilde{V}_{c}$
and $\epsilon_{\rm v}^{1} \approx 0.58\, \tilde{V}_{c}$.

Vacuum fluctuations from the Dirac sea thus shift and split the $n=0$ and $n=1$ levels; 
this is somewhat contrary to the common wisdom that filled levels can be regarded 
as dormant, with no active role to play.
The energy splitting 
\begin{equation}
\Delta \epsilon_{\rm v} \equiv 
\epsilon_{\rm v}^{0} - \epsilon_{\rm v}^{1} =\textstyle{1\over{8}}\, C\, \tilde{V}_{c}
\end{equation}
reflects the difference in their spatial distributions, 
as is clear from Eq.~(\ref{Vzeroreg}). 
The $n=0$ levels become higher in energy 
than the $n=1$ levels, when they are unoccupied.
Actually their relative positions vary with the filling of the $n=(0,1)$ sector.
To get a rough idea about how they vary, let us 
suppose filling the $n=1$ levels first (for $u= \mu_{\rm Z}=0$).
Equation~(\ref{VxDS}) then tells us to include extra contributions
$-|g^{01}|^2$  and $-|g^{11}|^2$ 
for $\epsilon_{\rm v}^{0}$ and $\epsilon_{\rm v}^{1}$, respectively;  
this yields $\epsilon_{\rm v}^{0} ={1\over{2}}\, (1- {1\over{2}}c_{1}^{2})\,  \tilde{V}_{c}$
and 
$\epsilon_{\rm v}^{1} =-{1\over{2}}\, (1- {1\over{2}}c_{1}^{2} - {1\over{4}} C)\,  \tilde{V}_{c}$.
If, instead, the $n=0$ levels were first filled, 
one would find 
$\epsilon_{\rm v}^{0} =-{1\over{2}}\, (1- {1\over{2}}c_{1}^{2})\, \tilde{V}_{c}$ and
$\epsilon_{\rm v}^{1} ={1\over{2}}\, (1- {1\over{2}}c_{1}^{2} - {1\over{4}} C)\, \tilde{V}_{c}$.
The ground-state energy is thus lower in the latter case.
This shows that the level structure indeed varies with filling of the sector 
and, in addition, indicates the unusual character of the orbital-polarized ground states, 
which we examine in detail later.

Let us next suppose that a pair of $n=0$ and  $n=1$ levels of the same spin or valley 
were filled (in the presence of some spin or valley breaking). 
One would then find that $\epsilon_{\rm v}^{0}$ and $\epsilon_{\rm v}^{1}$ 
reverse sign so that the filled $n=0$ level 
is now lower than the filled $n=1$ level;
this leads to a gap 
$2\epsilon_{\rm v}^{1}  =  (1+ {1\over{2}} c_{1}^{2} -\! {1\over{4}} C )\,  \tilde{V}_{c}$ 
between the filled and empty $n=1$ levels.
This (spin or valley) gap is larger than the possible $\lq\lq$orbital" gap
$ (1- {1\over{2}} c_{1}^{2} -\! {1\over{8}} C )\, \tilde{V}_{c}$ between 
the $n=1$ and $n=0$ levels, discussed above.
This suggests that relatively large energy gaps expected 
for the $\nu=\pm 2$ and $\nu=0$ ground states are either spin or valley gaps 
rather than orbital gaps, 
in accordance with Hund's rule and Ref. \onlinecite{BCNM}.

\section{HF calculation}

In this section we examine the level structure of the pseudo-zero-mode sector 
more quantitatively.
To this end, 
let us extract from $V_{X}$ in Eq.~(\ref{Vex}) the exchange interaction 
acting within the $n=(0,1)$ sector,   
\begin{eqnarray}
V_{\rm X}^{\rm pz}&=&
-  \sum_{\bf p}v_{\bf p}\gamma_{\bf p}^{2}\, \Omega,
\nonumber\\
\Omega &=&
\big\{ |g^{00}_{\bf p}|^{2}\, \nu^{00;ba}_{\beta\alpha} 
+ |g^{10}_{\bf p}|^{2}\, \nu^{11;ba}_{\beta\alpha} \big\}\, 
R^{00;ab}_{\alpha\beta;{\bf 0}}
\nonumber\\
&&+\big\{ |g^{10}_{\bf p}|^{2}\, \nu^{00;ba}_{\beta\alpha} 
+ |g^{11}_{\bf p}|^{2}\, \nu^{11;ba}_{\beta\alpha} \big\}\, 
R^{11;ab}_{\alpha\beta;{\bf 0}}
\nonumber\\
&&+ g^{00}_{\bf p}\, g^{11}_{\bf -p}\, 
\big\{ \nu^{10;ba}_{\beta\alpha}\, R^{01;ab}_{\alpha\beta;{\bf 0}} 
+  \nu^{01;ba}_{\beta\alpha}\,  R^{10;ab}_{\alpha\beta;{\bf 0}} \big\}.\ \ 
\label{Vxpz}
\end{eqnarray}
We consider
$H_{u} + V^{\rm DS}_{\rm X} + V_{\rm X}^{\rm pz}$, 
as the effective Hamiltonian, 
to explore the pseudo-zero-mode sector.
The exchange interaction $V^{\rm DS}_{\rm X} + V_{\rm X}^{\rm pz}$ 
is invariant under rotations in valley and spin, but not in orbital $(0,1)$ space.
In contrast, $H_{u}$, consisting of interlayer bias $u$ 
and the Zeeman energy $\mu_{\rm Z}$, 
can lift all of the valley, spin and orbital degeneracies.

The vacuum orbital splitting 
$\Delta \epsilon_{\rm v} ={1\over{8}}\, C\, \tilde{V}_{c}$, $u$ and $\mu_{\rm Z}$ 
would in general compete to determine 
how the pseudo-zero-mode octet is split in orbital, valley and spin 
when it is gradually filled with electrons.
In practice, a relatively weak bias $u$ can easily overtake  
the Zeeman energy
\begin{equation}
\mu_{\rm Z} \equiv g^{*}\mu_{B}B 
\approx 0.12\, B[{\rm T}]{\rm meV} = 1.4\,  B[{\rm T}]\, {\rm K};
\end{equation}
we allow $u$ to be arbitrary but keep $|u| < \tilde{V}_{c}$.
See Fig.~1, which depicts the empty $n=(0,1)$ sector (at $\nu=-4$) 
governed by $H_{u} + V^{\rm DS}_{\rm X}$,
with level spectra
\begin{equation}
\epsilon^{n \pm}_{\downarrow}
= \epsilon^{n}_{\rm v}  \pm \epsilon^{n}_{u} - \textstyle{1\over{2}}\mu_{\rm Z}, \ \
\epsilon^{n \pm}_{\uparrow} 
= \epsilon^{n}_{\rm v} \pm \epsilon^{n}_{u} + \textstyle{1\over{2}}\mu_{\rm Z} ,
\end{equation}
where $\epsilon^{na}_{\alpha}$ stands for the energy (per particle) 
of the $n_{a \alpha}$ level 
with orbital index $n\in (0,1)$, valley $a \in (+,-)$ and spin $\alpha \in (\uparrow, \downarrow)$.
For valley indices we use $(+,-)$ to specify  
that the (-)  state is lower in energy than the (+) state for each $n$ and spin $\alpha$;
accordingly, $(+,-) = (K,K')$ for $u>0$.
There are two possible level patterns, 
depending on (i) $\mu_{\rm Z}>(1-z) |u|$ or (ii) $(1-z)|u|>\mu_{\rm Z}$.
For definiteness, in this section we assume $\mu_{\rm Z} \gg |u|$ 
or even set $u\rightarrow 0$, and examine the hierarchy of broken-symmetry states
of case (i);
we discuss the general case with both $u$ and $\mu_{\rm Z}$ later.
In this setting, $H_{u}$ practically has little effect on the ground-state energies 
(since $\mu_{\rm Z}/V_{c}\ll 1)$.
Accordingly, and for further simplification,  we keep in mind 
that $H_{u}$ triggers spin breaking, suppress it and focus only on 
${\cal V}\equiv V^{\rm DS}_{\rm X}+V_{\rm X}^{\rm pz}$ in what follows.


\begin{figure}[tbp]
\includegraphics[scale=0.9]{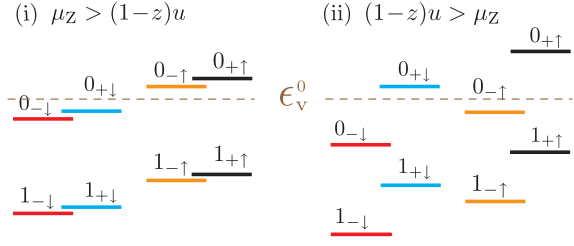}  
\caption{
Empty levels in the pseudo-zero-mode sector 
at $\nu=-4$;
for illustration, $(1-z)u/\mu_{\rm Z} = 0.2$ for (i) and 1.5 for (ii),
with
$\mu_{\rm Z}/\tilde{V}_{c}=0.05$. 
The dotted lines refer to the energy scale 
$\epsilon_{\rm v}^{0}= {1\over{2}}\,(1 + {1\over{2}}\, c_{1}^{2})\, \tilde{V}_{c}$. 
The valley index $\pm$ here stands for
$(+,-) =(K, K')$ with $u>0$.
}
\end{figure}


To diagonalize the exchange interaction ${\cal V}$ we proceed as follows:
Note again that ${\cal V}$ 
is invariant under rotations in spin and valley.
Spins and valleys, if driven externally, can therefore easily rotate.
Obviously, under the magnetic field $B_{z}=-B$, spins remain polarized in $(\uparrow, \downarrow)$.
Valleys may, however, rotate, $(K,K') \rightarrow (+,-)$. 
We suppose that the spin $(\uparrow, \downarrow)$ and  the rotated valley $(+,-)$
are good quantum numbers, and diagonalize each (valley, spin) sector via rotations 
in orbital $n=(0,1)$ space.  
We thus rotate 
$\psi^{n;a}_{\alpha} =(\psi_{\alpha}^{0;a}, \psi_{\alpha}^{1;a})$
first in valley space and then in orbital space,
\begin{equation}
\psi^{m;a}_{\alpha} (y_{0}) =  
[U(\hat{\theta}, \hat{\phi})]^{a}_{\ b}\,  [U(\theta, \phi)]^{m}_{\ n}\, \Phi^{n;b}_{\alpha} (y_{0}),
\label{rotatepsi}
\end{equation}
using the SU(2) matrix
\begin{equation}
U(\theta, \phi) =  \left(
\begin{array}{cc}
\cos (\theta/2) & -e^{ -i\phi}\sin (\theta/2) \\
e^{i \phi}\sin  (\theta/2) & \cos (\theta/2)\\
\end{array}
\right).
\end{equation}
By construction the orbital rotations $(\theta,\phi)$ refer to each (valley, spin)
sector, i.e., $(\theta,\phi)\rightarrow (\theta_{a \alpha},\phi_{a \alpha})$; 
for conciseness, we suppress this reference.
For the transformed fields $\Phi^{n;a}_{\alpha}$ we use $n=(0,1)$ and $a=(+,-)$.
For example, $\Phi^{0;+}_{\alpha}$ stands for $\psi^{0;K}_{\alpha}$ 
for $\theta=\hat{\theta}=0$
while it stands for $\psi^{1;K}_{\alpha}$ for $\theta=\pi$ and $\hat{\theta}=0$.
Actually, the valley rotations $(K,K') \rightarrow (+,-)$ are left undetermined 
for ${\cal V}$ alone.
They are to be fixed later (in Sec.~V) when we consider $H_{u}$ and a valley breaking 
due to interlayer Coulomb interactions.
Note next that $\phi$ and $\hat{\phi}$  can be absorbed into 
the relative phases between $\Phi^{0;a}_{\alpha}$ and $\Phi^{1;b}_{\alpha}$;  we therefore 
set $\phi=\hat{\phi}=0$ from now on.

Let $N^{0;ba}_{\beta\alpha}  \propto 
\langle G|{\Phi^{0;b}_{\beta}}^{\dag} \Phi_{\alpha}^{0;a}|G\rangle$
and $N^{1;ba}_{\beta\alpha} \propto 
\langle G|{\Phi_{\beta}^{1;b}}^{\dag} \Phi_{\alpha}^{1;a}|G\rangle$
denote the filling factors in the form of matrices in valley and spin.
By our assumption they are diagonal in both valley $a\in (+,-)$ 
and spin $\alpha \in (\uparrow,\downarrow)$.

Via rotations~(\ref{rotatepsi}), $V_{\rm X}^{\rm DS}$ reads
\begin{equation}
V^{\rm DS}_{\rm X}
= \epsilon_{\rm v}^{0} (\theta)\, {\cal R}^{00;aa}_{\alpha\alpha;{\bf 0}}
+ \epsilon_{\rm v}^{1} (\theta)\, {\cal R}^{11;aa}_{\alpha\alpha;{\bf 0}}
+ [\epsilon_{\rm v}^{0} (\theta)]'\, {\cal X}^{aa}_{\alpha\alpha},\ \ 
\label{VxDSrot}
\end{equation}
with
\begin{eqnarray}
\epsilon^{0}_{\rm v}(\theta) 
&=&\tilde{V}_{c}\, \big[\textstyle{1\over{2}}\,(1 + {1\over{2}}\, c_{1}^{2})
-{1\over{16}}\, C\, (1- \cos \theta) \big], 
\nonumber\\
\epsilon^{1}_{\rm v}(\theta) 
&=&\tilde{V}_{c}\, \big[\textstyle{1\over{2}}\,(1 + {1\over{2}}\, c_{1}^{2})
-{1\over{16}}\, C\, (1+ \cos \theta) \big].
\end{eqnarray}
Here ${\cal R}^{mn;ab}_{\alpha\beta;{\bf 0}}$ denote the charge operators 
for $\Phi^{n;b}_{\beta}$, i.e., 
$R^{mn;ab}_{\alpha\beta;{\bf 0}}$ with 
$\psi^{n;b}_{\beta}\rightarrow \Phi^{n;b}_{\beta}$,
and ${\cal X}^{ab}_{\alpha\beta}\equiv 
{\cal R}^{01;ab}_{\alpha\beta;{\bf 0}} 
+{\cal R}^{10;ab}_{\alpha\beta;{\bf 0}}$;
$[\cdots]' = (d/d\theta)[\cdots]$.
Similarly, $V_{\rm X}^{\rm pz}$ reads
\begin{eqnarray}
V_{\rm X}^{\rm pz} &=& \big[ E_{V}^{0}(\theta)\, N^{0;ba}_{\beta\alpha} 
+ \tilde{E}_{V}^{0}(\theta)\, N^{1;ba}_{\beta\alpha} \big] 
{\cal R}^{00;ab}_{\alpha\beta;{\bf 0}}
\nonumber\\
&& +\big[ E_{V}^{1}(\theta)\,  N^{0;ba}_{\beta\alpha} 
+ \tilde{E}_{V}^{1}(\theta)\,  N^{1;ba}_{\beta\alpha} \big] 
{\cal R}^{11;ab}_{\alpha\beta;{\bf 0}} 
\nonumber\\
&&+ \textstyle{1\over{2}}\, \big[ E_{V}^{0}(\theta)\, N^{0;ba}_{\beta\alpha}
- \tilde{E}_{V}^{1}(\theta) \, N^{1;ba}_{\beta\alpha} \big]'\, {\cal X}^{ab}_{\alpha\beta},
\label{transfVxpz}
\end{eqnarray}
where
\begin{eqnarray}
E_{V}^{0}(\theta)  
&=& -\tilde{V}_{c}\, \big[1 - \textstyle{1\over{16}}\, C (1- \cos \theta)^{2}\big] ,
\nonumber\\
\tilde{E}_{V}^{0}(\theta) &=& E_{V}^{1}(\theta) 
= -\tilde{V}_{c}\, \big[\textstyle{1\over{2}}c_{1}^{2} - {1\over{16}}\, C \sin^2 \theta \big],
\nonumber\\
\tilde{E}_{V}^{1}(\theta) &=& 
- \tilde{V}_{c}\, \big[1 - \textstyle{1\over{16}}\, C  (1+ \cos \theta)^{2} \big].
\end{eqnarray}
See Appendix A for details.
It is clear now that $V_{\rm X}^{\rm DS}$ and $V_{\rm X}^{\rm pz}$ are divided 
into four sectors, specified by $(a, \alpha)$, of the $2\times 2$
matrix Hamiltonians for $(\Phi_{\alpha}^{0;a}, \Phi_{\alpha}^{1;a})$.

Let us start filling the empty $n=(0,1)$ sector at $\nu=-4$, 
which consists of four $n=0$ and and four $n=1$  levels 
of energy 
$\epsilon^{0}_{\rm v}(0)={1\over{2}}\,(1 + {1\over{2}}\, c_{1}^{2})\, \tilde{V}_{c}$
and
$\epsilon^{1}_{\rm v}(0)=\epsilon^{0}_{\rm v}(0) - {1\over{8}}\, C\, \tilde{V}_{c}$;
numerically, $(\epsilon^{1}_{\rm v}(0), \epsilon^{0}_{\rm v}(0) ) = (0.58, 0.73)\, 
\tilde{V}_{c}$ at $B=10$T.  See Fig.~1.
There the orbital breaking $\Delta \epsilon_{\rm v}$ 
singles out the $\Phi^{1-}_{\downarrow}$ level (or $n_{a\alpha}=1_{-\downarrow}$)
as the lowest-lying one, which will thus be filled first. 
Suppose that this level is filled with fraction  $n_{1}\le 1$ 
and substitute $N^{1;--}_{\downarrow\downarrow}= n_{1}$ 
and $N^{0;--}_{\downarrow\downarrow}=0$ 
into ${\cal V}$.
The $(-,\downarrow)$ sector then consists of 
$\{ \epsilon^{1}_{\rm v}(\theta) + n_{1}\tilde{E}_{V}^{1}(\theta)\}\, 
{\cal R}^{11;- -}_{\downarrow\downarrow;{\bf 0}}$,
$\{ \epsilon^{0}_{\rm v}(\theta) + n_{1}\tilde{E}_{V}^{0}(\theta)\}\, 
{\cal R}^{00;- -}_{\downarrow\downarrow;{\bf 0}}$
and the $(0,1)$-mixed component 
\begin{equation}
-\big[\epsilon^{1}_{\rm v}(\theta) 
+ \textstyle{1\over{2}}\, n_{1}\tilde{E}_{V}^{1}(\theta)\big]'\, 
{\cal X}^{- -}_{\downarrow\downarrow}.
\end{equation}
Accordingly, on choosing $\theta$ so that 
this ${\cal X}^{- -}_{\downarrow\downarrow}$ term disappears, 
the $(-,\downarrow)$ sector is diagonalized.
Solving for $\theta$ then reveals that $\theta =0$ for $0\le n_{1} < 1/2$
while $\theta$ rises to $\pi/2$ as $n_{1}$ is increased 
from $1/2$ to 1; see Fig.~2~(a). 
We thus find that the filled $\Phi^{1-}_{\downarrow}$ 
and empty $\Phi^{0-}_{\downarrow}$ levels at $\nu=-3$
are equal mixtures $\propto \psi^{1;-}_{\downarrow} \pm \psi^{0;-}_{\downarrow}$ 
of the $n=0$ and $n=1$ orbitals, to be denoted as 
$1_{{\rm m} - \downarrow} \equiv 1_{- \downarrow}|_{\theta=\pi/2}$
and $0_{{\rm m} - \downarrow} \equiv 0_{- \downarrow}|_{\theta=\pi/2}$, 
respectively, in obvious notation.
The $\nu=-3$ ground state thus consists of the filled $1_{{\rm m}-\downarrow}$ 
and empty $0_{{\rm m} -\downarrow}$ levels of energy 
$\pm{1\over{2}}\,(1 - {1\over{2}}\, c_{1}^{2})\,  \tilde{V}_{c}$,
with an orbital gap 
$(1 - {1\over{2}}\, c_{1}^{2})\,  \tilde{V}_{c}$ [$\approx 0.55\,  \tilde{V}_{c}$ 
for $B=10$T].


\begin{figure}[tbp]
\includegraphics[scale=0.85]{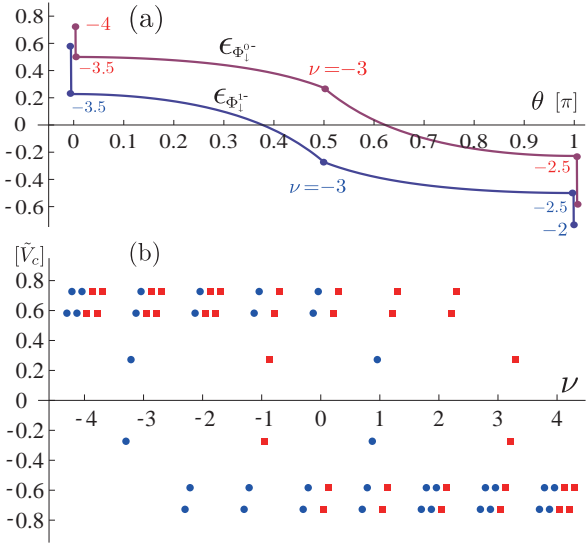}  
\caption{
(a) Variations of the spectra of 
the  $\Phi^{1-}_{\downarrow}$ and $\Phi^{0-}_{\downarrow}$ levels 
in the range $\nu \in [-4,-2]$. 
Note the symmetry 
$\epsilon_{\Phi^{0-}} (\theta)
= -\epsilon_{\Phi^{1-}} (\pi - \theta)$.
(b) Spectra of the pseudo-zero-mode Landau levels 
at each integer filling factor $\nu \in [-4,4]$, 
with $\mu_{\rm Z}> u>0$  and $\mu_{\rm Z}\rightarrow +0$.
Blue blobs refer to spin-down levels 
$(1_{-\downarrow}, 0_{-\downarrow}, 1_{+\downarrow},0_{+\downarrow})|_{\theta}$ 
from left to right and red squares refer to 
$(1_{-\uparrow}, 0_{-\uparrow}, 1_{+\uparrow},0_{+\uparrow}))|_{\theta}$;
empty levels have positive energy and occupied levels have negative energy.
}
\end{figure}


As seen from Fig.~2~(a), the $0_{-\downarrow}|_{\theta}$ level 
has come down along with $1_{-\downarrow}|_{\theta}$, 
owing to exchange interactions.
One can therefore reach the $\nu = -2$ state 
by filling the $0_{-\downarrow}|_{\theta}$ level,
i.e., by setting $N^{1;- -}_{\downarrow\downarrow}=1$ 
and  $N^{0;- -}_{\downarrow\downarrow}=n_{0}$ with $0\le n_{0} \le 1$.
Diagonalizing the $(-,\downarrow)$ sector
then shows that $\theta$ varies from $\pi/2$ to $\pi$ as $n_{0}$ is increased 
from 0 to $1/2$, 
and stays at $\pi$ until $n_{0}$ reaches 1. 
Note that, as one passes from $\nu=-4 \rightarrow -3 \rightarrow -2$,
the lowest energy level $\Phi^{1;-}_{\downarrow}$ has undergone the change 
$1_{-} \rightarrow 1_{{\rm m}-} (=1_{-}|_{\theta=\pi/2}) 
\rightarrow  0_{-} (= 1_{-}|_{\theta=\pi})$.
Here we notice the rule that each empty level characterized by 
$\Phi^{n; \pm}_{\alpha}|_{\theta=0}= \psi^{n; \pm}_{\alpha}$ 
turns into $\Phi^{n; \pm}_{\alpha}|_{\theta=\pi}$ 
when it is filled up. 
Note, in this connection, the relation
\begin{equation}
\epsilon^{1}_{\rm v}(\theta)
+ E^{1}_{V}(\theta) + \tilde{E}^{1}_{V}(\theta) 
= - \epsilon^{1}_{\rm v}(\theta)
\end{equation}
and an analogous one for $ \epsilon^{0}_{\rm v}(\theta)$, which 
imply  that the filled $\Phi^{1;-}_{\downarrow}$ 
and $\Phi^{0;-}_{\downarrow}$ levels have energies 
$(-\epsilon_{\rm v}^{1}(\pi), -\epsilon_{\rm v}^{0}(\pi) ) 
=( -\epsilon_{\rm v}^{0}(0), - \epsilon_{\rm v}^{1}(0) )$.
The $\nu=-2$ ground state thus consists of the filled $(-,\downarrow)$ sector
with $(\Phi^{1;-}, \Phi^{0;-})|_{\theta=\pi}=( -\psi^{0;-}, \psi^{1;-})$
of energy $( -\epsilon_{\rm v}^{0}(0), - \epsilon_{\rm v}^{1}(0) )$,
and is polarized in valley and spin, 
with a gap $2\epsilon_{\rm v}^{1}(0)$
 [$\approx 1.17\, \tilde{V}_{c}$ for $B=10$T].

Among the empty levels at $\nu=-2$ the lowest one is 
the $1_{+ \downarrow}$ level, 
rather than $1_{- \uparrow}$, for $\mu_{\rm Z}>u$.
To reach the $\nu=-1$ state, one therefore 
has to fill this $1_{+\downarrow}$ level.
The analysis is essentially the same as for the $\nu=-3$ case, and 
one finds the filled $1_{{\rm m} +\downarrow}$ and empty $0_{{\rm m} +\downarrow}$
levels of energy $\mp {1\over{2}}\,(1 - {1\over{2}}\, c_{1}^{2})\,  \tilde{V}_{c}$.
The $\nu=-1$ state is polarized in valley and orbital, 
with an orbital gap $(1 - {1\over{2}}\, c_{1}^{2})\,  \tilde{V}_{c}$.

One can further go to the $\nu=0$ state by filling the $0_{{\rm m} +\downarrow}$ level,
which thereby turns into the filled  $0_{+\downarrow}|_{\theta=\pi} =1_{+\downarrow}$ level.
The $\nu=0$ state is polarized in spin, 
with a gap $2\, \epsilon_{\rm v}^{1}(0)$.

Repeating essentially the same analysis for the spin-up sector
takes one to $0\le \nu \le 4$.
The resulting spectrum is electron-hole symmetric, as shown in Fig.~2~(b).

\section{interlayer Coulomb interaction}

The Coulomb potential between the two layers slightly differs  
from the intralayer potential $v_{\bf p}$, and leads to a weak valley-SU(2) breaking.
In this section we examine the effect of this breaking.

The relevant Coulomb interaction is written as
\begin{equation}
\tilde{V} = {1\over{2}}\sum_{\bf p}\,
( v^{+}_{\bf p}\rho_{\bf p}\rho_{\bf -p}
+ v^{-}_{\bf p}\rho^{-}_{\bf p}\rho^{-}_{\bf -p}),
\label{HCbr}
\end{equation}
where 
$v^{\pm}_{\bf p} = {1\over{2}}\, v_{\bf p}\,  (1\pm e^{-d\, |{\bf p}|})$
with the interlayer separation $d \approx 3.34 \AA$.
Here $\rho^{-}=\rho^{\rm upper} -\rho^{\rm lower}$ stands for 
the charge-density difference between the upper and lower layers.
One can write it as $\rho^{-}= \rho^{-;K} -\rho^{-;K'}$
with 
$\rho^{-;a}= \Psi^{a\dag}\Gamma \Psi^{a}$
for $a \in (K,K')$ and $\Gamma\equiv {\rm diag}(1,-1,-1,1)$.
Let us denote their Fourier transforms $\rho^{-;a}_{-{\bf p}} =\int d^{2}{\bf x}\,  
e^{i {\bf p\cdot x}}\,\rho^{-;a}$ as
\begin{equation}
(\rho^{-;K})_{-{\bf p}} = \gamma_{\bf p}\sum_{k,n,\alpha} h^{k n;K}_{\bf p}\, 
R^{k n;KK}_{\alpha\alpha;\bf p}, {\rm etc.},
\label{rhomk}
\end{equation}
where $h^{k n;K}_{\bf p} \equiv h^{k n}_{\bf p}|_{K}$ are given 
by $g^{k n}_{\bf p}|_{K}$ in Eq.~(\ref{gkn})
with the signs of the $b_{k}^{(2)}\, b_{n}^{(2)}$ and $b_{k}^{(3)}\, b_{n}^{(3)}$
terms reversed.
Within the $(0,1)$ sector
$h^{k n}_{\bf p}=g^{k n}_{\bf p}$ hold,
except for  
\begin{eqnarray}
h^{11}_{\bf p}&=&g^{11}_{\bf p}- \tilde{z}, 
\label{honeone} \\
\tilde{z} &\equiv& 2 \, c_{1}^{2}\, {(1- M^{2}z^{2})^{2}\over{\gamma'^{2}}}
= z_{c} - 2\, c_{1}^{2}z^{2}M^{2};
\end{eqnarray}
it is thus unnecessary to specify their valley indices.

Let us now substitute $v^{+}_{\bf p} = v_{\bf p}-v^{-}_{\bf p}$ into $\tilde{V}$, 
extract the $O(v^{-}_{\bf p})$ correction $\Delta V$ to $V$, 
and construct the HF Hamiltonian from it.
We first consider the Dirac-sea contribution
$\Delta V^{\rm DS}=\Delta V_{\rm D}^{\rm DS} + \Delta V_{\rm X}^{\rm DS}$.
The direct interaction reads
\begin{equation}
\Delta V_{\rm D}^{\rm DS}
= W \sum_{n \le -2} \delta h^{nn}_{\bf 0} \! 
\sum_{m=0,1}h^{mm}_{\bf 0}\,\delta R^{mm}_{\beta\beta;{\bf 0}}, 
\end{equation}
where 
$\delta h^{nn}_{\bf 0} \equiv h^{nn}_{\bf p= 0}|_{K} -h^{nn}_{\bf p=0}|_{K'}$, 
which is odd in $M \propto u/2$, and 
$\delta R^{m m}_{\beta\beta;{\bf 0}} \equiv R^{mm;KK}_{\beta\beta;{\bf 0}} 
-R^{mm;K'K'}_{\beta\beta;{\bf 0}}$;
\begin{equation}
W\equiv (d/\ell)\, V_{c}
\approx 0.0104\sqrt{B[{\rm T}]}\, \tilde{V}_{c}
\end{equation}
comes from $\rho_{0}v^{-}_{\bf p=0}= {1\over{2}}\, W$.
From $\Delta V_{\rm D}^{\rm DS}$ we have omitted  
a term $\propto W R^{mm;bb}_{\beta\beta;{\bf 0}}$ 
that shifts all levels uniformly.

This $\Delta V_{\rm D}^{\rm DS}$ leads to valley polarization 
$\propto h^{mm}_{\bf 0}\,\delta R^{mm}_{\beta\beta;{\bf 0}}$ 
of $O(M W)$ in the $n=(0,1)$ sector.
A direct numerical calculation shows that
\begin{equation}
\sum_{n \le -2} \delta h^{nn}_{\bf 0} \approx -16.0\,  M 
\label{deltaVD}
\end{equation}
for  $M\ll 1$ at $B=10$T.
$\Delta V_{\rm D}^{\rm DS}$ thus acts like $H_{u}$ and, when combined with it,
effectively reduces $u \rightarrow (1- 16\, W/\omega_{c})\, u$.
With this in mind, we henceforth understand 
that $H_{u}$ already involves the effect of $\Delta V_{\rm D}^{\rm DS}$.

The exchange interaction $\Delta V_{\rm X}^{\rm DS}$, 
on the other hand,      takes the form of $V^{\rm DS}_{\rm X}$ 
in Eq.~(\ref{VxDS})
with replacement $v_{\bf p}\rightarrow v^{-}_{\bf p}$ and 
$|g^{m'n;a}_{\bf p}|^{2}
 \rightarrow |h^{m'n;a}_{\bf p}|^{2}- |g^{m'n;a}_{\bf p}|^{2}$.
This $\Delta V_{\rm X}^{\rm DS}$ is calculable numerically.
Fortunately, for $M=0$, one can evaluate it exactly, 
as done for $V^{\rm DS}_{\rm X}$, with the result
\begin{equation}
\Delta V^{\rm DS}_{\rm X}
\stackrel{M\rightarrow 0}{=} 
{1\over{2}}\, \sum_{\bf p}v^{-}_{\bf p}\gamma_{\bf p}^{2}\,
 (|h^{11}_{\bf p}|^2 -|g^{11}_{\bf p}|^2)
R^{1 1;aa}_{\alpha\alpha;{\bf 0}},
\end{equation}
which, upon integration over ${\bf p}$, turns out to vanish to $O(W)$.
We have also verified this result numerically.

The main $O(W)$ corrections come from the interaction
$\Delta V^{\rm pz}$
acting within the $(0,1)$ sector.
See Appendix B for the details.
Here we quote only the result :
\begin{eqnarray}
\Delta V^{\rm pz}
&=& \textstyle{1\over{2}} W\, \{ 
- \nu \, {\cal R}^{nn;bb}_{\beta\beta;{\bf 0}} +I(\theta) 
+  t(\theta)\, \tilde{z}\, {\cal I}_{1}\}
\nonumber\\
&& + \textstyle{1\over{4}}  W \,\big[K(\theta) +  t'(\theta)\, \tilde{z}\,
 {\cal I}_{2} \big]- t(\hat{\theta})\, \Gamma_{W}+ \cdots,\
\nonumber\\
\Gamma_{W}&=& \{
\epsilon_{\rm cap}^{n}(\theta)\, \delta N^{0}_{\beta\alpha} 
+ \tilde{\epsilon}_{\rm cap}^{n}(\theta)\, \delta N^{1}_{\beta\alpha} \} \,
 \delta {\cal R}^{nn;}_{\alpha\beta;{\bf 0}}
\nonumber\\
&&+\textstyle{1\over{2}}\, 
 \{ \epsilon^{0}_{\rm cap} (\theta)\, \delta N^{0}_{\beta\alpha}\!
- \tilde{\epsilon}^{1}_{\rm cap}(\theta )\,\delta N^{1}_{\beta\alpha}\}'\, 
\delta {\cal X}_{\alpha\beta} 
\nonumber\\
&&+ W \{\Delta I(\theta) + \textstyle{1\over{2}}\Delta K(\theta) \},
\label{HW}
\end{eqnarray} 
with $n\in (0,1)$, $t(\theta) \equiv {1\over{2}}(\sin\theta)^2$ and 
\begin{eqnarray}
I(\theta) &=&
\big[ r_{0} \delta N^{0}_{\alpha\alpha} + r_{1}\delta N^{1}_{\alpha\alpha}\big]
\big[r_{0} \delta {\cal R}^{00}_{\beta\beta;{\bf 0}} 
+ r_{1}\delta {\cal R}^{11}_{\beta\beta;{\bf 0}} \big],
\nonumber\\
K(\theta) &=&\big[ (r_{0}^2)' \,  \delta N^{0}_{\alpha\alpha} 
- (r_{1}^2)' \, \delta N^{1}_{\alpha\alpha}\big] \delta {\cal X}_{\beta\beta}  
\nonumber\\
{\cal I}_{1}\!&=&\! \delta^{01}\! N^{++}_{\beta\alpha}\, 
\delta^{01} {\cal R}^{++}_{\alpha\beta;{\bf 0}}
+ \delta^{01}\! N^{--}_{\beta\alpha}\, \delta^{01} {\cal R}^{--}_{\alpha\beta;{\bf 0}},
\nonumber\\
{\cal I}_{2}&=& (\delta^{01} N^{++}_{\beta\alpha})\, {\cal X}^{++}_{\alpha\beta}
+ (\delta^{01} N^{--}_{\beta\alpha})\, {\cal X}^{--}_{\alpha\beta},
\label{Itheta}
\end{eqnarray}
where $r_{0} \equiv r_{0}(\theta) = 1\!-\! \tilde{z} s_{\theta}^2$ 
and $r_{1}\equiv r_{1}(\theta) = 1\!-\! \tilde{z} c_{\theta}^2$
with $s_{\theta} \equiv \sin(\theta/2)$ and $c_{\theta} \equiv \cos(\theta/2)$;
$r_{0}' = (d/d\theta)r_{0}(\theta)$, etc.
Here $\delta$ denotes the difference between the $(++)$ and $(--)$ components 
so that 
$\delta N^{0}_{\alpha\beta} \equiv N^{0;++}_{\alpha\beta} - N^{0;--}_{\alpha\beta}$, 
$\delta {\cal R}^{mn}_{\beta\beta;{\bf 0}} \equiv  {\cal R}^{mn;++}_{\beta\beta;{\bf 0}}
-{\cal R}^{mn;--}_{\beta\beta;{\bf 0}}$, 
$\delta {\cal X}_{\alpha\beta}\equiv 
{\cal X}^{++}_{\alpha\beta}- {\cal X}^{--}_{\alpha\beta}$, etc.
Similarly, $\delta^{01}$ denotes the $(0,1)$ difference: 
$\delta^{01}N^{++}_{\beta\alpha}
\equiv N^{00;++}_{\beta\alpha} -N^{11;++}_{\beta\alpha}$,
$\delta^{01}{\cal R}^{++}_{\alpha\beta;{\bf 0}} 
\equiv {\cal R}^{00;++}_{\alpha\beta;{\bf 0}}
-{\cal R}^{11;++}_{\alpha\beta;{\bf 0}}$, etc.
$\Delta I(\theta)$ and $\Delta K(\theta)$ stand for the crossed-spin variants of
$I(\theta)$ and $K(\theta)$ such that 
$\Delta K(\theta)
=[ \cdots]_{\uparrow\uparrow}\delta{\cal X}_{\downarrow\downarrow} 
+[ \cdots]_{\downarrow\downarrow}\delta{\cal X}_{\uparrow\uparrow}$,
etc.
From $\Delta V^{\rm pz}$ we have omitted terms $(\cdots) \propto t'(\hat{\theta})$ 
that eventually vanish 
when $t(\hat{\theta})$ is chosen to be stationary.

The valley-dependent term 
$- t(\hat{\theta})\, \Gamma_{W}$ in $\Delta V^{\rm pz}$ 
involves capacitance energies, defined as
\begin{eqnarray}
\epsilon_{\rm cap}^{0} (\theta)
&=&W\, \hat{D} (1-\cos \theta)^2, 
\nonumber\\
\epsilon_{\rm cap}^{1}  (\theta)
&=&\tilde{\epsilon}_{\rm cap}^{0} (\theta)
= W\, [ 1-c_{1}^{2} 
 -\tilde{z} + \hat{D}\,  (\sin\theta)^2],
\nonumber\\
\tilde{\epsilon}_{\rm cap}^{1}  (\theta)
&=&W\, \hat{D} (1+\cos \theta)^2,
\end{eqnarray}
where 
$\hat{D}= \textstyle{1\over{2}}\, c_{1}^{2}(1-c_{1}^{2})
- {1\over{2}}\tilde{z}+ {1\over{4}}\tilde{z}^2 
\stackrel{M\rightarrow 0}{=}  
- {1\over{4}}\, \tilde{z}+ {1\over{8}}\tilde{z}^2 <0$.
Here we encounter negative capacitance energies
\begin{equation}
\epsilon_{\rm cap}^{n} (\theta) < 0,\ \ \ \tilde{\epsilon}_{\rm cap}^{n} (\theta) < 0.
\end{equation}
This has an important consequence: 
It is easily seen that 
$\langle G|\Gamma_{W}|G\rangle <0$ for the possible ground states $|G\rangle$ 
discussed in the previous section.
The capacitance energy of the $n=(0,1)$ sector therefore is minimized 
for $t(\hat{\theta})=0$, 
or $\hat{\theta}=0,\pi$, which implies that 
there arise essentially no valley rotations in the pseudo-zero-mode sector 
in bilayer graphene.  
This is in sharp contrast to conventional quantum Hall (QH) systems, 
for which positive capacitance energies favor valley-symmetric and 
antisymmetric combinations ($\hat{\theta}=\pi/2$) for eigenstates. 
Normally both direct and exchange interactions contribute to the capacitance energy 
and their contributions tend to cancel 
for zero layer separation $d\rightarrow 0$ or to $O(W)$;
this leaves positive $O(d^2/\ell^2)$ capacitance energies 
for conventional systems. 

These negative capacitance energies for the $n=(0,1)$ sector 
are traced back to the fact that 
the $n=1$ modes, in particular, are distributed on both layers 
with ratio 1 to $\sqrt{\bar{z}/2}$ in amplitude,
or with ratio $1-\tilde{z}/2$ to $\tilde{z}/2$ in charge, 
as seen from Eqs.~(\ref{beone}) and ~(\ref{honeone}). 
The direct Coulomb interaction thereby gets somewhat weaker $(r_{0}, r_{1} <1)$
and is overtaken by the exchange contribution, leaving  $(\epsilon_{\rm cap}^{n},
\tilde{\epsilon}_{\rm cap}^{n})$ negative to $O(W)$.
This fact, though clear in terms of 4-component electron spinors $\Psi^{K}$ and 
$\Psi^{K'}$, 
could easily be overlooked 
when one uses the approximate 2-component spinor formalism~\cite{MF} 
for the description of bilayer graphene.

We study the effect of this $O(W)$ valley breaking as a perturbation 
to the eigenstates found in Sec.~IV.
One may set $t(\hat{\theta})=0$ and 
read from Eq.~(\ref{HW}) the $O(W)$ corrections to the spectrum:
\begin{eqnarray}
E_{W\downarrow}^{1\pm}\! &=&\! \textstyle{1\over{2}} W 
\big\{ 
- \nu  \pm  r_{0}r_{1}\, \delta N^{0}_{\alpha\alpha} 
\pm  r_{1}^{2}\, \delta N^{1}_{\alpha\alpha}
- \tilde{z}\, t(\theta)\, \delta^{01}\! N^{\pm \pm}_{\downarrow\downarrow}
\big\},
\nonumber\\
E_{W\downarrow}^{0 \pm}\! &=&\! \textstyle{1\over{2}} W  
\big\{
-\nu 
 \pm r_{0}^{2}\, \delta N^{0}_{\alpha\alpha} 
\pm  r_{0}r_{1}\, \delta N^{1}_{\alpha\alpha}
+ \tilde{z}\, t(\theta)\, \delta^{01}\! N^{\pm \pm}_{\downarrow\downarrow} 
\big\};
\nonumber\\
\end{eqnarray}
analogously for the spin-up sector.
Here $E_{W\downarrow}^{0-}$, 
e.g., stands for the correction to 
the ${\cal R}^{00;- -}_{\downarrow\downarrow;{\bf 0}}$ sector.
For the $\nu=-3$ state 
with only one filled level, e.g.,
one finds
\begin{eqnarray}
&&E_{W\downarrow}^{1-} \stackrel{\theta=\pi/2}{=} 
( 2 - \textstyle{1\over{4}}\tilde{z}+{1\over{8}} \tilde{z}^2)\, W,
\nonumber\\
&&E_{W\downarrow}^{0-} 
\stackrel{\theta=\pi/2}{=} 
( 2 - \textstyle{3\over{4}}\tilde{z} + {1\over{8}} \tilde{z}^2)\, W,
\nonumber\\
&&E_{W\downarrow}^{1+} \stackrel{\theta= 0}{=} 
(1 +\tilde{z}- \textstyle{1\over{2}} \tilde{z}^2)\, W,
\nonumber\\
&&E_{W\downarrow}^{0+} \stackrel{\theta= 0}{=} 
( 1 + \textstyle{1\over{2}}\tilde{z})\, W.
\end{eqnarray}

\section{General case} 

We have so far supposed that $\mu_{\rm Z}\gg u \sim 0$. 
Let us now relax it and
consider the full effective Hamiltonian 
\begin{equation}
H_{\rm eff}={\cal V} + H_{u} +\Delta {\cal V},
\end{equation}
with  ${\cal V}=V^{\rm DS}_{\rm X}+V_{\rm X}^{\rm pz}$ and
$\Delta {\cal V}=\Delta V^{\rm DS}_{\rm D}+ \Delta V^{\rm DS}_{\rm X} 
+ \Delta V^{\rm pz}$,
to explore the pseudo-zero-mode sector with 
both $\mu_{\rm Z}$ and bias $u$.
We keep $|u| \ll V_{c}$
so that we can still use the $u=0$ expressions for ${\cal V} + \Delta{\cal V}$,
with $u$ retained  only in $H_{u}$ as a small perturbation. 
[We have in mind $u \sim$ a few meV at $B\sim$ 10 T. ] 
Actually this is a reasonable approximation if one notes the following:
Possible $O(u)$ valley breakings that
come from the Dirac-sea contribution $(V^{\rm DS}_{\rm X} +\cdots)$ 
can effectively be taken care of 
by rescaling $u$ in $H_{u}$, as we have seen for $\Delta V_{\rm D}^{\rm DS}$
in Eq.~(\ref{deltaVD}).
In addition, $V^{\rm pz} + \Delta V^{\rm pz}$, acting within the $(0,1)$ sector, 
contains no $O(M)$ corrections; note in this connection 
that $z_{c}$ and $\tilde{z}$ differ from $z$ only to $O(z^2M^2)$.

The energy scale $\sim V_{c}$ of the $n=(0,1)$ sector
is set by ${\cal V}$, 
and relatively weak (spin, valley and orbital) breakings in 
$H_{\rm eff}$ trigger symmetry breaking of this sector, leading 
to exchange-enhanced gaps 
at each integer filling factor $\nu \in [-4,4]$. 
We have seen that negative capacitance energies suppress valley rotations.
Accordingly,  one can assign $(+,-)=(K, K')$ or $\hat{\theta}=0$,
assuming, without loss of generality, $u\ge 0$.
Via rotations~(\ref{rotatepsi}), $H_{u}$ somewhat changes its form; 
the main change is 
$\epsilon_{u}^{n}\, \delta R^{nn}_{\beta\beta;{\bf 0}} 
\rightarrow 
\epsilon_{u}^{n}(\theta)\, \delta {\cal R}^{nn}_{\beta\beta;{\bf 0}}
+ \cdots$,
with  
\begin{eqnarray}
\epsilon^{0}_{u} (\theta) &=& \textstyle{1\over{2}}\, u \cos \hat{\theta} 
\{ 1- {1\over{2}}\, z\,(1-\cos \theta) \},
\nonumber\\
\epsilon^{1}_{u} (\theta)  
&=& \textstyle{1\over{2}}\, u \cos \hat{\theta} 
\{ 1- {1\over{2}}\, z\,(1+\cos \theta) \}.
\end{eqnarray}


\begin{figure}[tbp]
\includegraphics[scale=0.85]{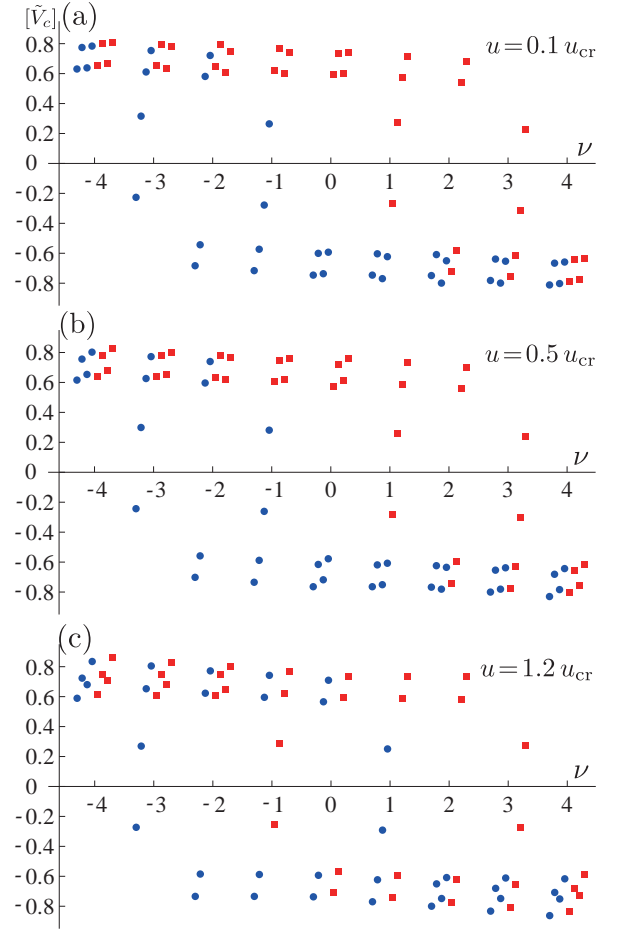}  
\caption{
Spectra of the pseudo-zero-mode sector 
for each integer filling factor $\nu \in [-4,4]$,
with $\mu_{\rm Z}/\tilde{V}_{c}  \sim 0.027$ taken as a typical value at $B=10$ T;
this yields $u_{\rm cr}/\tilde{V}_{c} \approx 0.093$.
(a) $u=0.1\, u_{\rm cr}$. (b) $u=0.5\,  u_{\rm cr}$. 
(c) $u=1.2\, u_{\rm cr}$. 
Blobs and squares denote levels ordered as in Fig.~2 (b).
}
\end{figure}


Let us look at Fig.~1 again, which depicts the empty $n=(0,1)$ sector
for the two cases,
(i) $(1-z) u < \mu_{\rm Z}$ and (ii) $(1-z) u > \mu_{\rm Z}$;
see also Fig.~3.
There the lowest-lying level is  $1_{-\downarrow}$ in both cases and is filled first.
As the $1_{-\downarrow}$ level is being filled, 
it comes down in energy, followed by the $0_{-\downarrow}$ level
(of the same valley and spin) paired  via the exchange interaction. 
The $\nu=-3$ state will therefore consist of the filled 
$1_{{\rm m}-\downarrow}$
and empty $0_{{\rm m} -\downarrow}$ levels
of energy
\begin{eqnarray}
\epsilon^{1-}_{\downarrow}&=&
-\textstyle{1\over{2}}(1- \textstyle{1\over{2}}c_{1}^{2})\,  \tilde{V}_{c} 
+ [\epsilon^{1}_{u} (\theta) + E^{1-}_{W\downarrow}]_{\theta=\pi/2},
 \nonumber\\
\epsilon^{0-}_{\downarrow}&=&
+\textstyle{1\over{2}}(1- \textstyle{1\over{2}}c_{1}^{2})\,  \tilde{V}_{c} 
+ [\epsilon^{0}_{u} (\theta) + E^{0-}_{W\downarrow}]_{\theta=\pi/2}, 
\end{eqnarray}
which lead to an orbital gap,
\begin{equation}
\epsilon^{\rm gap}_{\nu=-3} = (1- \textstyle{1\over{2}}c_{1}^{2})\,  \tilde{V}_{c} 
- {1\over{2}}\tilde{z}W.
\label{gapnuthree}
\end{equation}

As a result, the $\nu=-2$ state is also unique and consists of the filled  
$(1_{-\downarrow}, 0_{-\downarrow})|_{\theta=\pi}$ levels, polarized in both valley and spin.
(If, instead, an empty $n=1$ level were to come down in passing to $\nu=2$, 
it would lead to an orbital-polarized 
$\nu=-2$ state, which is energetically not favored, as noted at the end of Sec.~III. 
We therefore exclude this possibility.)

Among the empty levels at $\nu=-2$ the lowest-lying level 
is either $1_{-\uparrow}$ or $1_{+\downarrow}$, depending on case (i) and (ii).
Comparing their energy levels reveals 
that $1_{-\uparrow}$ is lower for $u<u_{\rm cr}$, 
and vice versa, with the critical value $u_{\rm cr}$ given by
\begin{equation}
  (1-z)\, u_{\rm cr} =\mu_{\rm Z} + \lambda(\tilde{z})\, W,
\end{equation}
where $\lambda(\tilde{z}) \equiv 2-3\tilde{z} +\tilde{z}^2$.
Cases (i) and (ii) are thus distinguished by this more precise criterion
$u<u_{\rm cr}$ and $u>u_{\rm cr}$.
Correspondingly, the $\nu=-2$ level gap differs in structure for the two cases,
\begin{eqnarray}
\epsilon^{\rm gap}_{\nu= -2}|_{\rm (i)} &=& 
2\, \epsilon^{1}_{\rm v}(0) + (1-z)\, u- \lambda(\tilde{z})\, W,
\nonumber\\
\epsilon^{\rm gap}_{\nu= -2}|_{\rm (ii)} &=& 
2\, \epsilon^{1}_{\rm v}(0)  + \mu_{\rm Z}.
\label{Egapnutwo}
\end{eqnarray}
The $\nu=-2$ gap changes from a valley gap to a spin gap 
as $u$ is increased across $u_{\rm cr}$.

Numerically, at $B=10$T, $\lambda(\tilde{z})\, W/\tilde{V}_{c} \approx 0.049$
and $\mu_{\rm Z}/\tilde{V}_{c}  \sim 0.027$ 
for typical values $\mu_{\rm Z} \sim1.2\,  {\rm meV}$ and $\epsilon_{b} \sim 5$;
this leads to $u_{\rm cr} \sim 0.093\, \tilde{V}_{c} \sim 4$ meV;
or more generally,  $u_{\rm cr}/\tilde{V}_{c} \sim (0.065, 0.13, 0.17)$
for $B=(5,20,30)$ T.  
Note that the Coulombic interlayer valley breaking $\lambda(\tilde{z})W$
is practically comparable to $\mu_{\rm Z}$.

Filling either $1_{+\downarrow}$ or $1_{-\uparrow}$
leads to the $\nu=-1$ state.
It consists of the filled $1_{{\rm m} +\downarrow}$ level 
(and the paired $0_{{\rm m} +\downarrow}$)
[over the filled $\nu=-2$ basis  $(1_{-\downarrow},0_{-\downarrow})|_{\theta=\pi}$] 
for $u<u_{\rm cr}$ while it consists of the filled $1_{{\rm m} -\uparrow}$ level 
(and the paired $0_{{\rm m} -\uparrow}$) for $u>u_{\rm cr}$.
The resulting level gap $\epsilon^{\rm gap}_{\nu=-1}$, nevertheless, 
turns out to be the same as $\epsilon^{\rm gap}_{\nu=-3}$.

The $\nu=0$ state is reached by filling the paired $n=1$ level, and 
differs in composition, depending on $u$; see Figs.~3 (b) and 3 (c).
It is spin-polarized for $u<u_{\rm cr}$
while it is valley-polarized for $u>u_{\rm cr}$, 
with a level gap
\begin{eqnarray}
\epsilon^{\rm gap}_{\nu= 0}|_{\rm (i)} &=&  
2\, \epsilon^{1}_{\rm v}(0)  + \mu_{\rm Z} - (1-z)\, u,
\nonumber\\
\epsilon^{\rm gap}_{\nu= 0}|_{\rm (ii)} 
&=&  2\, \epsilon^{1}_{\rm v}(0) +(1 \! -\! z)\, u
- \mu_{\rm Z}  - 2\lambda(\tilde{z})\, W.
\label{Egapnuzero}
\end{eqnarray}

It will be clear now how to reach the $\nu=1\sim 4$ states.
One eventually finds that the pseudo-zero-mode octet has
perfectly particle-hole symmetric spectra, as shown in Fig.~3
for each integer filling factor $\nu \in [-4, 4]$.
It is evident from the figures how each level behaves as the $n=(0,1)$
sector is gradually filled, and that the spectra change in pattern 
for $u< u_{\rm cr}$ and for $u > u_{\rm cr}$.

It turns out that the level gaps at odd integer filling 
are all purely orbital gaps
of the same magnitude, 
$\epsilon^{\rm gap}_{\nu=\pm 3}= \epsilon^{\rm gap}_{\nu=\pm 1}$,
which are relatively small 
and insensitive to both $\mu_{\rm Z}$ and $u$. 
It is interesting to examine experimentally how these possible gaps respond 
to a tilted magnetic field.
An additional parallel magnetic field $B_{\parallel}$ effectively works 
to increase $\mu_{\rm Z}$ and $u_{\rm cr}$.
Equation~(\ref{gapnuthree}) therefore suggests 
that the $\nu=\pm1,\pm3$ gaps are insensitive to $B_{\parallel}$.

So far we have supposed filling the $n=(0,1)$ sector gradually for a fixed $u$.
When $u$ is applied for a fixed density or $\nu$, 
cases (i) and (ii) are simply connected for the $\nu=\pm 3, \pm 2$ states 
but not for the $\nu=0, \pm 1$ states.
We discuss this point in the next section.

\section{comparison with experiments}

Experimentally full splitting of 
the pseudo-zero-mode Landau levels has been observed~\cite{FMY,ZCZJ,WAFM,MFWAY,VJB}
at high magnetic fields.
The quantum numbers  (especially, valley and orbital ones) 
of the resulting broken-symmetry states remain unknown yet, but
the way they emerge with magnetic field $B$
reveals the relative magnitude of the associated energy gaps,  
\begin{equation}
\epsilon^{\rm gap}_{\nu= 0} > \epsilon^{\rm gap}_{\nu= \pm 2} 
\gg \epsilon^{\rm gap}_{\nu= \pm 1,\pm 3}.
\end{equation}
In particular, Feldman {\it et al,}\cite{FMY}, via conductance measurements 
in suspended bilayer graphene,
observed full degeneracy lifting for $B\ge 3$T, and noted that 
the $\nu=0$, $\nu=-2$ and $\nu=-1$ states become apparent at 0.1 T,  0.7 T and 
at 2.7 T, respectively.
Similarly, in bilayer devices on SiO$_{2}$/Si substrates
Zhao {\it et al.}\cite{ZCZJ}
observed full degeneracy lifting for $B\ge$ 20 T 
and noted that the resistance minima 
for $\nu=-2$ and $\nu= -3$ are barely affected 
by a parallel magnetic field $B_{\parallel}$.
Both experiments indicate that the $\nu=-1$ and $\nu=-3$ QH states emerge 
at similar magnetic fields.
These observed features are consistent with our picture of the pseudo-zero-mode octet 
for case (i) $\mu_{\rm Z}>u\sim 0$. 

The energy gaps  inferred or extracted from experiments~\cite{FMY,ZCZJ,WAFM,MFWAY,VJB} 
are generally smaller that those
expected from $\tilde{V}_{c} \approx (70/\epsilon_{b})\, \sqrt{B[{\rm T}]}$ meV.
In particular, Martin {\it et al.},\cite{MFWAY} via local compressibility measurements of 
suspended bilayer graphene devices, probed energy gaps of size
$\epsilon^{\rm gap}_{\nu= 0}\approx 1.7\,  B[{\rm T}]$ meV
and $\epsilon^{\rm gap}_{\nu= \pm 2}\approx 1.2\,  B[{\rm T}]$ meV   
(and also $\epsilon^{\rm gap}_{\nu= \pm1}\sim 0.1\,  B[{\rm T}]$ with less data points);
also a recent transport experiment~\cite{VJB} reports a somewhat larger gap 
$\epsilon^{\rm gap}_{\nu= 0}\approx 5.5\,  B[{\rm T}]$ meV.
Such linear $B$ scaling of gaps may appear to contradict the naive $\sqrt{B}$ behavior of 
$\tilde{V}_{c}$.
A possible resolution is that charge is efficiently screened\cite{KSbgr} due to quantum fluctuations 
of the valence band;  $\epsilon_{b}$ is more enhanced for smaller $B$.
This screening weakens $\tilde{V}_{c}$ and 
could entail quasi-linear $B$ behavior of $\tilde{V}_{c}$.

Let us try to handle some numbers.
Supposing that $\epsilon^{\rm gap}_{\nu= 0} \sim 17$ meV at 10T  
(using the numbers quoted above)  and 
putting it into Eq.~(\ref{Egapnuzero}) yields $\tilde{V}_{c} \sim 14$ meV or  $\epsilon_{b} \sim 15$,
 i.e., quite a suppression of $\tilde{V}_{c}$.
 On the other hand,  according to Eqs.~(\ref{Egapnutwo}) and~(\ref{Egapnuzero}), 
$\epsilon^{\rm gap}_{\nu= 0}- \epsilon^{\rm gap}_{\nu= -2} 
\stackrel{u=0}{=} \mu_{\rm Z} + \lambda (\tilde{z}) W > \mu_{\rm Z}$.
%
Supposing $\epsilon^{\rm gap}_{\nu= 0}- \epsilon^{\rm gap}_{\nu= -2} \sim 5$ meV 
at $B=$10T then yields $\epsilon_{b} \sim 3$.  
This suggests that the difference $\epsilon^{\rm gap}_{\nu= 0}- \epsilon^{\rm gap}_{\nu= -2}$ contains a significant Coulombic contribution other than the Zeeman energy
(though the effect of charge screening is not clear). 
This is a mere order-of-magnitude estimate and 
one has to remember that 
experimental numbers themselves have considerable uncertainties;
the value of $\epsilon_{b}$ easily changes with the input 
one uses.
At least, the observed values of energy gaps indicate 
that they are due to the Coulomb interaction, 
which, though screened, still sets a far larger scale than 
the spin and valley breakings.

The effect of an electric field $E_{\perp}$ on broken-symmetry states 
has also been studied~\cite{WAFM, VJB} for suspended bilayer graphene.
Weitz {\it et al.}~\cite{WAFM} measured 
the two-terminal conductance as a function of density and electric field
at various magnetic fields.
They find, in particular, that the $\nu=0$ state shows quantized conductance 
except near two large values of $E_{\perp}$, 
and interpret this as suggesting the crossover 
of the spin-polarized $\nu=0$ state at low $E_{\perp}$ to
layer-polarized $\nu=0$ states at large $E_{\perp}$. 
The data also suggest a similar crossover 
at large $E_{\perp}$ for the $\nu=\pm 1$ states.


\begin{figure}[tbp]
\includegraphics[scale=0.9]{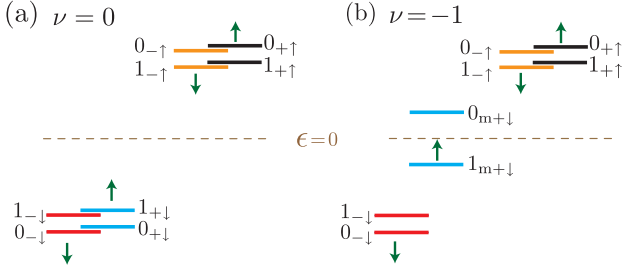}  
\caption{
The pseudo-zero-mode sector (a) at $\nu=0$ and 
(b) at $\nu=-1$. Green arrows indicate how levels with valley $(+,-)$ 
move with increasing interlayer bias $u$. 
}
\end{figure}


The crossover phenomena are seen in our picture as well. 
In our picture the $\nu=0$ state is different in composition 
for $u\sim0$ and $u\gg u_{\rm cr}$.
The crossover of the spin-polarized $\nu=0$ state to the valley-polarized state,
when $E_{\perp}$ is applied for a fixed density, 
has to proceed by changing their composition.
As illustrated in Fig.~4 (a),
the filled $(0_{+\downarrow}, 1_{+\downarrow})$ levels go up and 
the empty $(0_{-\uparrow}, 1_{-\uparrow})$ levels come down with increasing $u$.
A crossover thus takes place when the gap is closed,
$\epsilon^{0+}_{\downarrow} \sim 0$,
 i.e., around\cite{fntwo} $u \sim 2\epsilon^{1}_{\rm v}(0)$; 
 conductance quantization is naturally lost around the zero gap.
An observed value  $E_{\perp} \sim 50$ mV/nm for $B=2.65$T, 
read from Ref.~\onlinecite{WAFM}, e.g., 
amounts to $u\sim 17$ meV, which indeed is
on the order of magnitude of the Coulombic gap.
A similar crossover around $u \sim 2\epsilon^{1}_{\rm v}(0)$ is also expected 
for the $\nu=-1$ state; see Fig.~4 (b).

\section{Summary and discussion}

In a magnetic field bilayer graphene has an octet of zero-energy Landau levels
due  to an extra twofold degeneracy in Landau orbitals $n=0$ and $n=1$;
this degeneracy is not accidental and has a topological origin.
These levels evolve, in the presence of spin and valley breakings and Coulomb interactions,
into pseudo-zero-mode Landau levels.

In this paper we have studied the structure of this pseudo-zero-mode octet 
and, in particular, pointed out that its orbital degeneracy is lifted 
by Coulombic quantum fluctuations of the Dirac sea (the valence band).
This quantum effect derives from the Dirac-sea contribution 
to the electron selfenergy, and the $n=0$ and $n=1$ modes 
are split due to their different spatial distributions. 
It is analogous to the Lamb shift,~\cite{Lambshift} 
which is the energy difference between the $2P_{1/2}$ and $2S_{1/2}$ states 
of the hydrogen atom due to vacuum fluctuations.

For bilayer graphene vacuum fluctuations make the $n=0$ modes higher in energy 
than the $n=1$ modes (of the same spin and valley) when they are empty, 
but the order is reversed when they are filled.   
We have seen that such vacuum effects are intimately correlated 
with the Coulomb interaction within the zero-mode sector 
so that the spectra of this special sector correctly realize
the particle-hole symmetry of the basic Hamiltonian.
In a sense, a multi-component version of the Lamb shift emerges 
in bilayer graphene owing to spin and valley degrees of freedom.
It will be clear now that this quantum effect, 
though simply overlooked in earlier approaches 
that rely on Coulomb interactions projected to the zero-mode sector alone, 
has to be properly taken into account in studying the dynamics 
in the pseudo-zero-mode sector.

Another finding of the paper is negative capacitance energies, 
which suppress possible valley rotations in the pseudo-zero-mode sector. 
They come from the nontrivial structure of the interlayer Coulomb interaction 
in bilayer graphene, such that the $n=1$ modes are distributed 
on both layers while the $n=0$ modes are localized on either layer;
the valley and layer degrees of freedom are generally not the same 
for bilayer graphene.

Experimentally it will be a challenge to study the structure 
of the pseudo-zero-mode levels directly 
via measurements of cyclotron resonance. 
Cyclotron resonance obeys the selection rule~\cite{AFcr} $\Delta |n| =\pm 1$ 
and is diagonal in spin and valley.
There is no cyclotron resonance within the pseudo-zero-mode sector 
at $\nu=0$ and $\nu=\pm 2$ because of a mismatch in spin or valley.
Cyclotron resonances in the $1\leftrightarrow 0$ channel will emerge 
around $\nu=\pm 1$ and $\nu=\pm 3$ 
with resonance energies
 [ranging from $\Delta\epsilon_{\rm v}$ to $(1- {1\over{2}}c_{1}^{2})\tilde{V}_{c}$]
considerably higher than previously expected.~\cite{BCNM} 
Such resonances are in practice rather difficult to detect 
because the pseudo-zero-modes carry current much smaller 
(by factor $z M=z\, (u/2)/\omega_{c}\ll 1$) than other $|n|\ge 2$ modes,
as read from the Hamiltonian 
\begin{equation}
H_{A} \sim - { e\ell \, \omega_{c}\over{\sqrt{2}}}\, zMc_{1} 
 \Big( A^{\dag} \delta R^{01}_{\alpha\alpha; {\bf 0}}
  + A \delta R^{10}_{\alpha\alpha; {\bf 0}} \Big)
\end{equation}
for a time-varying potential $A=A_{x}(t) +i A_{y}(t)$,
where 
$\delta R^{01}_{\alpha\alpha;{\bf 0}} 
= R^{01;KK}_{\alpha\alpha;{\bf 0}} - R^{01;K'K'}_{\alpha\alpha;{\bf 0}}$. 
One can equally well explore the structure of the pseudo-zero-mode sector 
by looking into the $n=-2 \rightarrow 1$ and  $1 \rightarrow 2$ channels
of cyclotron resonance at different filling factors $\nu \in [-4,4]$.

\acknowledgments

This work was supported in part by a Grant-in-Aid for Scientific Research
from the Ministry of Education, Science, Sports and Culture of Japan 
(Grant No. 21540265).

\appendix

\section{rotations in valley and $(0,1)$ space}

In this appendix we outline the derivation of the exchange interaction 
$V_{\rm X}^{\rm pz}$ in Eq.~(\ref{transfVxpz}). 
Let us start with  
$V_{\rm X}^{\rm pz}=-  \sum_{\bf p}v_{\bf p}\gamma_{\bf p}^{2}\,\Omega$ 
in Eq.~(\ref{Vxpz}). 
Via rotations~(\ref{rotatepsi}),  the kernel $\Omega$ is rewritten as
\begin{eqnarray}
\Omega &=&\big[ f_{0}(\theta)\, N^{0;ba}_{\beta\alpha} 
+ \tilde{f}_{0}(\theta)\, N^{1;ba}_{\beta\alpha} \big] 
{\cal R}^{00;ab}_{\alpha\beta;{\bf 0}} 
\nonumber\\
&&+\big[ f_{1}(\theta)\,  N^{0;ba}_{\beta\alpha} 
+ \tilde{f}_{1}(\theta)\,  N^{1;ba}_{\beta\alpha} \big] 
{\cal R}^{11;ab}_{\alpha\beta;{\bf 0}} 
\nonumber\\
&&+ \textstyle{1\over{2}}\, \big[ f'_{0}(\theta)\, N^{0;ba}_{\beta\alpha}
- \tilde{f}'_{1}(\theta) \, N^{1;ba}_{\beta\alpha} \big]\,
{\cal X}^{ab}_{\alpha\beta},
\label{basicform}
\end{eqnarray}
where 
 $N^{n;ba}_{\beta\alpha}  \propto 
 \langle G|{\Phi^{n;b}_{\beta}}^{\dag} \Phi_{\alpha}^{n;a}|G\rangle$
denote the filling-factor matrices of the level $n\in (0,1)$,
${\cal R}^{mn;ab}_{\alpha\beta;{\bf 0}}$ 
are the charge operators for the transformed fields,  
i.e., $R^{mn;ab}_{\alpha\beta;{\bf 0}}$ with 
$\psi^{n;b}_{\beta}\rightarrow \Phi^{n;b}_{\beta}$,
and ${\cal X}^{ab}_{\alpha\beta}\equiv e^{-i\phi}
{\cal R}^{01;ab}_{\alpha\beta;{\bf 0}} 
+e^{i\phi}{\cal R}^{10;ab}_{\alpha\beta;{\bf 0}}$.
The coefficient functions are defined as
\begin{eqnarray}
f_{0}(\theta) &=& |g^{00}_{\bf p}|^2  
+ s_{\theta}^{4}\, (|g^{11}_{\bf p}|^2 - |g^{00}_{\bf p}|^2)
+ 2 s_{\theta}^2 c_{\theta}^2\, \Xi,
\nonumber\\
&=& 1 - s_{\theta}^{4}\, ( c_{1}^{2}q^2 
- \textstyle{1\over{4}}c_{1}^{4}q^4) + 2 s_{\theta}^2 c_{\theta}^2\, \Xi,
\nonumber\\
f_{1}(\theta) &=&  \tilde{f}_{0}(\theta)
=|g^{10}_{\bf p}|^2  
+ s_{\theta}^{2}c_{\theta}^2\, (|g^{11}_{\bf p}|^2 - |g^{00}_{\bf p}|^2 - 2 \, \Xi),
\nonumber\\
&=&  \textstyle{1\over{2}}c_{1}^{2}q^2 - s_{\theta}^{2} c_{\theta}^2\, 
( c_{1}^{2}q^2 - {1\over{4}}c_{1}^{4}q^4 +2\, \Xi),
\nonumber\\
\tilde{f}_{1}(\theta) &=&  |g^{00}_{\bf p}|^2 
+ c_{\theta}^4\,  (|g^{11}_{\bf p}|^2 - |g^{00}_{\bf p}|^2)
+ 2 s_{\theta}^2 c_{\theta}^2\, \Xi, 
\nonumber\\
\Xi &=& g^{00}_{\bf p}g^{11}_{\bf p} 
+ |g^{10}_{\bf p}|^2 -|g^{00}_{\bf p}|^2\label{ftheta},
\end{eqnarray}
where $q \equiv \ell |{\bf p}|$, $s_{\theta} \equiv \sin(\theta/2)$ and 
$c_{\theta} \equiv \cos (\theta/2)$ for short.
The special combination $\Xi$, on substituting $g^{mn}_{\bf p}$, turns out to vanish. 
Noting the formula 
$\sum_{\bf p}v_{\bf p}\gamma_{\bf p}^2 \,[1, q^2, q^4] 
= [1,1,3]\sqrt{\pi/2}\, V_{c}$
and carrying out integrals over ${\bf p}$ 
with functions $f_{n}(\theta)$ and $\tilde{f}_{n}(\theta)$  
leads to $V_{\rm X}^{\rm pz}$ in Eq.~(\ref{transfVxpz}).

\section{ $O(v^{-}_{\bf p})$ correction $\Delta V^{\rm pz}$}

In this appendix we construct
the $O(v^{-}_{\bf p}) \sim O(V_{c}\, d/\ell)$ correction $\Delta V^{\rm pz}$
to the Coulomb interaction $V^{\rm pz}$ acting within the $n=(0,1)$ sector.
The $O(d/\ell)$ direct interaction reads
\begin{eqnarray}
\Delta V_{\rm D}^{\rm pz}
&=&  -\textstyle{1\over{2}} W\,
\nu^{nn;aa}_{\alpha\alpha}\, R^{mm;bb}_{\beta\beta;{\bf 0}}
\nonumber\\
&&+ \textstyle{1\over{2}} W\,
(h^{nn}_{\bf 0}\,  \delta \nu^{nn}_{\alpha\alpha})\, 
h^{mm}_{\bf 0}\,\delta R^{mm}_{\beta\beta;{\bf 0}} 
\end{eqnarray}
with $n,m \in (0,1)$; 
$ \delta \nu^{nn}_{\alpha\beta} \equiv  \nu^{nn;KK}_{\alpha\beta} 
- \nu^{nn;K'K'}_{\alpha\beta}$
and $W \equiv 2\rho_{0}v^{-}_{\bf p=0}=(d/\ell)\, V_{c}$.
This, via rotations~(\ref{rotatepsi}), is rewritten as
\begin{eqnarray}
\Delta V_{\rm D}^{\rm pz}
&=& -\textstyle{1\over{2}}\, \nu\,W\,  {\cal R}^{mm;bb}_{\beta\beta;{\bf 0}} 
\nonumber\\
&&+ \{ \textstyle{1\over{2}} -t(\hat{\theta}) \}\, W 
\{ I(\theta) + {1\over{2}} K(\theta) \} + \cdots,
\end{eqnarray}
with $I(\theta)$ and $K(\theta)$ defined in Eq.~(\ref{Itheta});
$t(\hat{\theta}) = {1\over{2}} (\sin \hat{\theta})^2$. 
The first term is proportional to the filling of the $n=(0,1)$ sector 
and is here adjusted to vanish for $\nu=0$.
The suppressed terms $(\cdots) \propto t'(\hat{\theta})$ eventually vanish 
when $\hat{\theta}$ is chosen to make 
$t(\hat{\theta})$ stationary;
we omit such terms from now on.

The exchange interaction $\Delta V^{\rm pz}_{\rm X}$ 
inherits the structure of
$V_{\rm X}^{\rm pz}$ in Eq.~(\ref{Vxpz}).
It consists of two parts:
$\Delta V_{{\rm X}+}^{\rm pz} =\sum_{\bf p}v^{-}_{\bf p}\gamma_{\bf p}^2\, 
(\Omega+ \Omega|_{g\rightarrow h})$
for the $KK' +K'K$ combination and 
$\Delta V_{{\rm X}-}^{\rm pz} =\sum_{\bf p}v^{-}_{\bf p}\gamma_{\bf p}^2\, 
(\Omega -\Omega|_{g\rightarrow h})$
for the $KK +K'K'$ combination,
where $\Omega|_{g\rightarrow h}$ stands for $\Omega$ 
in Eq.~(\ref{basicform}) with $g^{mn}_{\bf p}$ replaced 
by $h^{mn}_{\bf p}$.
For calculations one may  use Eq.~(\ref{ftheta}) and note that 
$\Xi|_{g\rightarrow h} = h^{11}_{\bf p}- g^{11}_{\bf p}= -\tilde{z}$,
together with 
\begin{equation}
\sum_{\bf p}v^{-}_{\bf p}\gamma_{\bf p}^2 [1, q^2, q^4] =  (W/2)\, [1,2,8] 
+ O(d^2/\ell^2).
\end{equation}
The result for $\Delta V_{{\rm X}+}^{\rm pz}$ to $O(W)$ is 
\begin{eqnarray}
\Delta V_{{\rm X}+}^{\rm pz} \!\!
&=&
t(\hat{\theta}) W \Big[ (F^{n}_{V} \delta N^{0}_{\beta\alpha}
+ \tilde{F}^{n}_{V} \delta N^{1}_{\beta\alpha})\, 
\delta {\cal R}^{nn}_{\alpha\beta;{\bf 0}}
\nonumber\\
&&+\textstyle{1\over{2}} \{F^{0}_{V}(\theta) 
\delta N^{0}_{\beta\alpha}\!
- \tilde{F}^{1}_{V}(\theta) \delta N^{1}_{\beta\alpha} \}' \, 
\delta {\cal X}_{\alpha\beta} \Big], \ \ \
\end{eqnarray}
with $n\in (0,1)$; $\delta {\cal R}^{n'n'}_{\alpha\beta;{\bf 0}} 
\equiv {\cal R}^{n'n';++}_{\alpha\beta;{\bf 0}}
- {\cal R}^{n'n';--}_{\alpha\beta;{\bf 0}}$,
$\delta {\cal X}_{\alpha\beta}\equiv
{\cal X}^{++}_{\alpha\beta}-
{\cal X}^{--}_{\alpha\beta}$, etc.
Here
\begin{eqnarray}
F^{0}_{V}(\theta) &=& 1 - D\, (1- \cos \theta)^{2}
- \textstyle{1\over{4}}\tilde{z}\, \sin^{2}\theta,
\nonumber\\ 
F^{1}_{V}(\theta) &=&\tilde{F}^{0}_{V} (\theta)
= c_{1}^{2} -  (D - \textstyle{1\over{4}}\tilde{z})\, \sin^2 \theta,
\nonumber\\ 
\tilde{F}^{1}_{V}(\theta) 
&=& 1 -  D\, (1+ \cos \theta)^{2}- \textstyle{1\over{4}}\tilde{z}\, \sin^{2}\theta,
\end{eqnarray} 
with $D ={1\over{2}}\, c_{1}^{2}(1-c_{1}^{2})$.
 
Similarly one finds
\begin{eqnarray}
\Delta V_{{\rm X}-}^{\rm pz}
&=&\textstyle{1\over{2}} \tilde{z}\, W \, 
\{ t(\theta)\, I_{1} +{1\over{2}}\,  t'(\theta)\, I_{2}\},  
\nonumber\\
I_{1} &=& {\cal I}_{1}
-t(\hat{\theta}) \{ \delta N^{0}_{\beta\alpha} - \delta N^{1}_{\beta\alpha} \}
(\delta {\cal R}^{00}_{\alpha\beta; {\bf 0}} 
-\delta {\cal R}^{11}_{\alpha\beta; {\bf 0}}),
\nonumber\\
I_{2}&=& {\cal I}_{2}
-t(\hat{\theta}) (\delta N^{0}_{\beta\alpha} - \delta N^{1}_{\beta\alpha} )\, 
\delta {\cal X}_{\alpha\beta},
\end{eqnarray}
with ${\cal I}_{1}$ and ${\cal I}_{2}$ defined in Eq.~(\ref{Itheta}).
Collecting $\Delta V_{\rm D}^{\rm pz}$, $\Delta V_{{\rm X}+}^{\rm pz}$ and 
$\Delta V_{{\rm X}-}^{\rm pz}$ 
and combining terms involving $t(\hat{\theta})$ 
into capacitance energies yields Eq.~(\ref{HW}).



\begin{thebibliography}{99}
\bibitem{ehasymmetry} 
For the electron-hole ($eh$) asymmetry in bilayer graphene 
see Refs.~\onlinecite{KScr}, \onlinecite{Malard} and \onlinecite{ehASref}.
In the present paper we focus on the $eh$ symmetric case, with  
the $eh$ asymmetry left for future study.
\bibitem{fntwo} This is only a rough estimate. 
For a quantitative analysis one has to take into account 
the response of the Dirac sea to a large bias $u$.
\end{thebibliography}
\end{document}